\documentclass[aps,prd,12pt,preprint,superscriptaddress,nofootinbib,floatfix]{revtex4}
\usepackage{latexsym}
\usepackage{amsfonts}
\usepackage[latin1]{inputenc}
\usepackage{subfig}
\usepackage{graphicx}
\usepackage{mathrsfs}
\usepackage{amssymb}
\usepackage{amsmath}
\usepackage{verbatim}
\begin{document}

\begin{flushleft}
{SHEP-10-16}\\
\today
\end{flushleft}
\vspace*{-0.25cm}

\title{A Renormalisation Group Equation Study of the Scalar Sector \\[0.15cm]
of the Minimal $B-L$ Extension of the Standard Model}
\vspace*{1.0cm}
\author{Lorenzo Basso}
\affiliation{
School of Physics \& Astronomy, University of Southampton,\\
Highfield, Southampton SO17 1BJ, UK}
\affiliation{
Particle Physics Department, Rutherford Appleton Laboratory, \\Chilton,
Didcot, Oxon OX11 0QX, UK}
\author{Stefano Moretti}
\affiliation{
School of Physics \& Astronomy, University of Southampton,\\
Highfield, Southampton SO17 1BJ, UK}
\affiliation{
Particle Physics Department, Rutherford Appleton Laboratory, \\Chilton,
Didcot, Oxon OX11 0QX, UK}
\author{Giovanni Marco Pruna}
\affiliation{
School of Physics \& Astronomy, University of Southampton,\\
Highfield, Southampton SO17 1BJ, UK}
\affiliation{
Particle Physics Department, Rutherford Appleton Laboratory, \\Chilton,
Didcot, Oxon OX11 0QX, UK}

\begin{abstract}
{\small \noindent
We present the complete set of Renormalisation Group Equations (RGEs) at one loop for the non-exotic minimal 
$U(1)$ extension of the Standard Model (SM). It includes all models that are anomaly-free with the SM fermion 
content augmented by one Right-Handed (RH) neutrino per generation. We then pursue the numerical study of the pure 
$B-L$ model, deriving the triviality and vacuum stability bounds on an enlarged scalar sector comprising 
one additional Higgs singlet field with respect to the SM.
}
\end{abstract}
\maketitle

\newpage
\section{Introduction}
The Large Hadron Collider has essentially been built to confirm or disprove the existence of
one or more Higgs bosons. A lot of effort has therefore been put into studying models that can accommodate the 
Higgs mechanism of Electroweak Symmetry Breaking (EWSB), of which the Higgs (pseudo)scalar particles are remnants. 
Among these, the most studied one is the Standard Model (SM). Unfortunately, the SM is flawed. There is now 
experimental evidence of new phenomena that cannot be explained by the SM, notably (very small) neutrino masses. 
At the same time, it should be noted that the accidental $U(1)_{B-L}$ global symmetry (where $B(L)$ is the Barion(Lepton) number) is not anomalous in the SM 
with massless neutrinos, but its origin is not understood. It thus becomes appealing to extend the SM to explain 
simultaneously the existence of both neutrino masses and the $B-L$ global symmetry by gauging the $U(1)_{B-L}$ 
group and subject it to spontaneous EWSB induced by the Higgs mechanism (therefore generating a 
massive $Z'$ 
state on the same footing as massive $W^\pm$ and $Z$ states are generated from the breaking of the
$SU(2)_L\times U(1)_Y$ gauge symmetry of the SM). Consequently, this requires that the fermion and 
scalar spectra are enlarged to account for gauge anomaly cancellations in such a way as to evade direct 
searches.
 Minimally, this requires the addition of a scalar 
singlet and three (massive) right-handed neutrinos, one per generation \cite{Jenkins:1987ue,Buchmuller:1991ce,Khalil:2006yi,B-L}, 
the latter entering the see-saw mechanism to explain the smallness of the 
detected (SM-like) neutrino masses \cite{see-saw}. 

Generally, $U(1)$ gauge factors mix and the mixing is controlled by further gauge couplings \cite{Aguila-Coughlan}. These extra parameters can then be reabsorbed in an effective parameterisation \cite{CPW}. In the model we are considering, two Abelian groups are present, and just one extra coupling is therefore needed to account for the mixing, which can effectively be reabsorbed. The arising model is a minimal (i.e., one-dimensional) $U(1)$ extension of the SM, spanning over several benchmark models
among which the ``pure'' $B-L$ model \cite{Khalil:2006yi,B-L,bbms} is a particularly simple example, as we will describe in the next Section.

Unsurprisingly, because of the simple nature of such a model,  
following the experimental results on neutrino masses, a plethora of papers have been published 
studying the phenomenology of the $B-L$ model at 
colliders. They have dealt with the detectabilty of the $Z'$ boson (see \cite{Jenkins:1987ue} for
earlier studies on generic hadron colliders)  at the LHC \cite{bbms,Emam:2007dy,Mine,disc_pot,LH09} 
and at a future Linear Collider (LC) \cite{Freitas:2004hq,bbmp,LC09}, some analyses concentrating on the $Z'$ 
decaying 
via heavy neutrinos, in particular into three \cite{bbms} and four \cite{khalil-4l} leptons in the final state,  
with distinctive displaced vertices due to long lived neutrinos, a clear signature of physics beyond the SM. 
Also, the testability at the LHC of the see-saw mechanism in this model has been evaluated in detail 
\cite{Perez:2009mu}.

In comparison to the gauge and neutrino sectors, the Higgs part of this model has undergone much less scrutiny.
Apart from the benchmark study of \cite{Emam:2007dy} dating back a few years, only recently 
a systematic analysis of the Higgs sector of the $B-L$ model has started, in the attempt to define
the boundaries of the associated parameter space. Ref.~\cite{B-L-unitarity} dealt with the limits 
stemming from the imposition of perturbative unitarity on the model. Here, we intend to pursue further
into this attempt, by investigating the triviality and vacuum stability conditions ensuing in the $B-L$ model,
through a RGE analysis aiming at defining the physical values of the masses and couplings of the two Higgs 
states emerging in the model after EWSB, the latter depending upon the maximum energy scale after 
which also such a scenario ceases to be valid and further new physics dynamics ought to be invoked. We believe 
that, with the LHC now on line, it is of paramount importance to theoretically constrain the Higgs sector of a
new physics scenario that, while incorporating the SM, it remedies its major flaw without leading to a 
proliferation of new particles and/or interactions, thereby retaining much of the predictivity and testability
of the SM.

In the past and yet recent years, a lot of effort has been spent for similar studies. For reviews on the SM and on some of its extensions, see Refs.~\cite{sher,Ellis:2007wa} and references therein. Concerning the study presented here, earlier works focusing on extra singlet scalars or $E_6$-inspired $U(1)$ augmented gauge groups in non-suspersimmetric \cite{Binoth:1996au,Gonderinger:2009jp} and suspersimmetric \cite{Saxena:2007xx,Athron:2009bs} extensions of the SM, respectively, have already been considered.

The plan of the paper is as follows. In section \ref{sect:parameter} we describe the model under study.
In section~\ref{sect:comp_det} we describe our computational techniques. The following section~\ref{sect:results}
presents our numerical results while we conclude in section \ref{sect:summa}. We also have an appendix, where
we list the RGEs of the model that we have dealt with.

\section{The parameterisation}\label{sect:parameter}
The model under study is the minimal $U(1)_{B-L}$ extension of the SM (see ref.~\cite{B-L} for conventions and references), 
in which the SM gauge group is augmented by a $U(1)$ factor, related to the Baryon minus Lepton ($B-L$) 
gauged number. In the complete model, the classical gauge invariant Lagrangian,
obeying the $SU(3)_C\times SU(2)_L\times U(1)_Y\times U(1)_{B-L}$
gauge symmetry, can be decomposed as:
\begin{equation}\label{L}
\mathscr{L}=\mathscr{L}_s + \mathscr{L}_{YM} + \mathscr{L}_f + \mathscr{L}_Y \, .
\end{equation}

The scalar Lagrangian is:
\begin{equation}\label{new-scalar_L}
\mathscr{L}_s=\left( D^{\mu} H\right) ^{\dagger} D_{\mu}H + 
\left( D^{\mu} \chi\right) ^{\dagger} D_{\mu}\chi - V(H,\chi ) \, ,
\end{equation}
with the scalar potential given by
\begin{eqnarray}\nonumber
V(H,\chi )&=&m^2H^{\dagger}H + \mu ^2\mid\chi\mid ^2 +\left( \begin{array}{cc} H^{\dagger}H& \mid\chi\mid ^2\end{array}\right)
					\left( \begin{array}{cc} \lambda _1 & \frac{\lambda _3}{2} \\
			 \frac{\lambda _3}{2} & \lambda _2 \\ \end{array} \right) \left( \begin{array}{c} H^{\dagger}H \\ \mid\chi\mid ^2 \\ \end{array} \right)\\
			  \nonumber \\ \label{BL-potential}
		&=&m^2H^{\dagger}H + \mu ^2\mid\chi\mid ^2 + \lambda _1 (H^{\dagger}H)^2 +\lambda _2 \mid\chi\mid ^4 + \lambda _3 H^{\dagger}H\mid\chi\mid ^2  \, ,
\end{eqnarray}
where $H$ and $\chi$ are the complex scalar Higgs doublet and singlet fields, respectively.

We generalise the SM discussion of spontaneous EWSB to the more complicated classical potential of eq.
(\ref{BL-potential}). To determine the condition for $V(H,\chi )$ to be bounded from below, it is sufficient to study its behaviour
for large field values, controlled by the matrix in the first line of eq. (\ref{BL-potential}). Requiring such a matrix to be 
positive-definite, we obtain the conditions:
\begin{equation}\label{inf_limitated}
4 \lambda _1 \lambda _2 - \lambda _3^2>0 \, ,
\end{equation}
\begin{equation}\label{positivity}
\lambda _1, \lambda _2 > 0 \, .
\end{equation}
If the above conditions are satisfied, we can proceed to the minimisation of $V$ as a function of constant 
Vacuum Expectation Values (VEVs) for the two Higgs
fields. Making use of gauge invariance, it is not restrictive to assume:
\begin{equation}\label{min}
\left< H \right> \equiv \left( \begin{array}{c} 0 \\ \frac{v}{\sqrt{2}} \end{array} \right)\, , 
	\hspace{2cm} \left< \chi \right> \equiv \frac{x}{\sqrt{2}}\, ,
\end{equation} 
with $v$ and $x$ real and non-negative. The physically most interesting solutions to the minimisation of eq.~(\ref{BL-potential}) are obtained for $v$ and $x$ both non-vanishing:
\begin{eqnarray}\label{sol_min1}
v^2 &=& \frac{-\lambda _2 m^2 + \frac{\lambda _3}{2}\mu ^2}{\lambda _1 \lambda _2 - \frac{\lambda _3^{\phantom{o}2}}{4}} \, ,\\
\nonumber  \\ \label{sol_min2}
x^2 &=& \frac{-\lambda _1 \mu ^2 + \frac{\lambda _3}{2}m ^2}{\lambda _1 \lambda _2 - \frac{\lambda _3^{\phantom{o}2}}{4}} \, .
\end{eqnarray}

To compute the scalar masses, we must expand the potential in eq. (\ref{BL-potential}) around the minima
in eqs. (\ref{sol_min1}) and (\ref{sol_min2}).
We denote by
$h_1$ and $h_2$ the scalar fields of definite masses, $m_{h_1}$ and $m_{h_2}$ respectively, and we conventionally choose
$m^2_{h_1} < m^2_{h_2}$. After standard manipulations, the explicit expressions for the scalar mass eigenvalues and eigenvectors are:
\begin{eqnarray}\label{mh1}
m^2_{h_1} &=& \lambda _1 v^2 + \lambda _2 x^2 - \sqrt{(\lambda _1 v^2 - \lambda _2 x^2)^2 + (\lambda _3 xv)^2} \, ,\\ \label{mh2}
m^2_{h_2} &=& \lambda _1 v^2 + \lambda _2 x^2 + \sqrt{(\lambda _1 v^2 - \lambda _2 x^2)^2 + (\lambda _3 xv)^2} \, ,
\end{eqnarray}
\begin{equation}\label{scalari_autostati_massa}
\left( \begin{array}{c} h_1\\h_2\end{array}\right) = \left( \begin{array}{cc} \cos{\alpha}&-\sin{\alpha}\\ \sin{\alpha}&\cos{\alpha}
	\end{array}\right) \left( \begin{array}{c} h\\h'\end{array}\right) \, ,
\end{equation}
where $-\frac{\pi}{2}\leq \alpha \leq \frac{\pi}{2}$ fulfils\footnote{In all generality, the whole interval $0\leq \alpha < 2\pi$ is halved
because an orthogonal transformation is invariant under $\alpha \rightarrow \alpha + \pi$. We could re-halve the interval by noting
that it is invariant also under $\alpha \rightarrow -\alpha$ if we permit the eigenvalues inversion, but this is forbidden by our
convention $m^2_{h_1} < m^2_{h_2}$. Thus $\alpha$ and $-\alpha$ are independent solutions.}:\label{scalar_angle}
\begin{eqnarray}\label{sin2a}
\sin{2\alpha} &=& \frac{\lambda _3 xv}{\sqrt{(\lambda _1 v^2 - \lambda _2 x^2)^2 + (\lambda _3 xv)^2}} \, ,\\ \label{cos2a}
\cos{2\alpha} &=& \frac{\lambda _1 v^2 - \lambda _2 x^2}{\sqrt{(\lambda _1 v^2 - \lambda _2 x^2)^2 + (\lambda _3 xv)^2}}\, .
\end{eqnarray}

For our numerical study of the extended Higgs sector, it is useful to invert eqs.~(\ref{mh1}), (\ref{mh2}) and (\ref{sin2a}), 
to extract the parameters in the Lagrangian in terms of the physical quantities $m_{h_1}$, $m_{h_2}$ and $\sin{2\alpha}$:
\begin{eqnarray}\nonumber
\lambda _1 &=& \frac{m_{h_2}^2}{4v^2}(1-\cos{2\alpha}) + \frac{m_{h_1}^2}{4v^2}(1+\cos{2\alpha}),\\ \nonumber
\lambda _2 &=& \frac{m_{h_1}^2}{4x^2}(1-\cos{2\alpha}) + \frac{m_{h_2}^2}{4x^2}(1+\cos{2\alpha}),\\ \label{inversion}
\lambda _3 &=& \sin{2\alpha} \left( \frac{m_{h_2}^2-m_{h_1}^2}{2xv} \right).
\end{eqnarray}

Moving to the $\mathscr{L}_{YM}$, the non-Abelian field strengths therein are the same as in the SM whereas the Abelian ones can be written as follows:
\begin{equation}\label{La}
\mathscr{L}^{\rm Abel}_{YM} = 
-\frac{1}{4}F^{\mu\nu}F_{\mu\nu}-\frac{1}{4}F^{\prime\mu\nu}F^\prime _{\mu\nu}\, ,
\end{equation}
where
\begin{eqnarray}\label{new-fs3}
F_{\mu\nu}		&=&	\partial _{\mu}B_{\nu} - \partial _{\nu}B_{\mu} \, , \\ \label{new-fs4}
F^\prime_{\mu\nu}	&=&	\partial _{\mu}B^\prime_{\nu} - \partial _{\nu}B^\prime_{\mu} \, .
\end{eqnarray}
In this field basis, the covariant derivative is:
\begin{equation}\label{cov_der}
D_{\mu}\equiv \partial _{\mu} + ig_S T^{\alpha}G_{\mu}^{\phantom{o}\alpha} 
+ igT^aW_{\mu}^{\phantom{o}a} +ig_1YB_{\mu} +i(\widetilde{g}Y + g_1'Y_{B-L})B'_{\mu}\, .
\end{equation}

To determine the gauge boson spectrum, we have to expand the scalar kinetic terms as for the SM. We expect
that there exists a massless gauge boson, the photon, whilst the other gauge bosons become massive. The extension we are studying
is in the Abelian sector of the SM gauge group, so that the charged gauge bosons $W^\pm$ will have masses given by their SM expressions,
being related to the $SU(2)_L$ factor only. Using the unitary-gauge parameterisation, 
 the kinetic terms in eq.
(\ref{new-scalar_L}) become:%
\begin{eqnarray}\nonumber
\left( D^{\mu} H\right) ^\dagger D_{\mu}H &=& \frac{1}{2}\partial ^{\mu} h \partial _{\mu}h + \frac{1}{8} (h+v)^2 \big(
				0\; 1 \big) \Big[ g W_a ^{\phantom{o}\mu }\sigma _a + g_1B^{\mu}+\widetilde g B'^{\mu}
				\Big] ^2 \left( \begin{array}{c} 0\\1\end{array} \right) \\ \nonumber
	&=& \frac{1}{2}\partial ^{\mu} h \partial _{\mu}h + \frac{1}{8} (h+v)^2 \left[ g^2 \left| W_1 ^{\phantom{o}\mu } -
		iW_2 ^{\phantom{o}\mu } \right| ^2  \right.\\ \label{boson_masses1}
	&&	\left. \hspace{4cm} + \left( gW_3 ^{\phantom{o}\mu } - g_1 B^{\mu} - \widetilde g B'^{\mu}\right) ^2 \right] \, ,
\end{eqnarray}
and
\begin{eqnarray}\label{boson_masses2}
\left( D^{\mu} \chi\right) ^\dagger D_{\mu}\chi &=& \frac{1}{2}\partial ^{\mu} h' \partial _{\mu}h' + \frac{1}{2}(h'+x)^2
(g_1' 2B'^{\mu})^2\, ,
\end{eqnarray}
where we have taken $Y^{B-L}_\chi = 2$ in order to guarantee the gauge invariance of the Yukawa terms (see eq. (\ref{L_Yukawa})).
In eq.~(\ref{boson_masses1}) we can recognise immediately the SM
charged gauge bosons $W^\pm$, with $\displaystyle M_W=gv/2$ as in the SM. The other gauge boson masses are
not so simple to identify, because of mixing. In fact, in analogy with the SM, the fields of definite mass are linear combinations
of $B^\mu$, $W_3^\mu$ and $B'^\mu$. The explicit expressions are:
\begin{equation}\label{neutral_bosons}
\left( \begin{array}{c} B^{\mu} \\ W_3^{\phantom{o}\mu}\\ B'^{\mu} \end{array}\right) = \left(
		\begin{array}{ccc}
		\cos{\vartheta _w} & -\sin{\vartheta _w}\cos{\vartheta '} & \sin{\vartheta _w}\sin{\vartheta '}\\
		\sin{\vartheta _w} & \cos{\vartheta _w}\cos{\vartheta '} & -\cos{\vartheta _w}\sin{\vartheta '}\\
		0 & \sin{\vartheta '} & \cos{\vartheta '}
		\end{array} \right) \left( \begin{array}{c} A^{\mu} \\ Z^{\mu}\\ Z'^{\mu} \end{array}\right)\, ,\\
\end{equation}
with $-\frac{\pi}{4}\leq \vartheta '\leq \frac{\pi}{4}$, such that:
\begin{equation}\label{tan2theta_prime}
\tan{2\vartheta '}=\frac{2\widetilde{g}\sqrt{g^2+g_1^2}}{\widetilde{g}^2 + 16(\frac{x}{v})^2 g_1^{'2}-g^2-g_1^2}
\end{equation}
and
\begin{eqnarray}\nonumber
M_A &=& 0\, ,\\ \label{Mzz'}
M_{Z,Z'} &=& \sqrt{g^2 + g_1^2}\cdot \frac{v}{2} \left[ \frac{1}{2}\left(  \frac{\widetilde{g}^2 + 16(\frac{x}{v})^2 g_1^{'2}}{g^2 + g_1^2}+1
	\right) \mp \frac{\widetilde{g}}{\sin{2\vartheta '}\sqrt{g^2+g_1^2}} \right] ^{\frac{1}{2}}\, ,
\end{eqnarray}
where
\begin{displaymath}
\sin{2\vartheta '}=\frac{2\widetilde{g}\sqrt{g^2+g_1^2}}{\sqrt{\left(\widetilde{g}^2 + 16(\frac{x}{v})^2 g_1^{'2}-g^2-g_1^2\right) ^2
		+(2\widetilde{g})^2(g^2+g_1^2)}}\, .
\end{displaymath}
LEP experiments \cite{Abreu:1994ria} constrain $|\vartheta '| \lesssim 10^{-3}$. Present constraints on the VEV $x$ (see section~\ref{sect:exp-lim}) allow a generous range of $\widetilde{g}$.

The fermionic Lagrangian (where $k$ is the
generation index) is given by
\begin{eqnarray} \nonumber
\mathscr{L}_f &=& \sum _{k=1}^3 \Big( i\overline {q_{kL}} \gamma _{\mu}D^{\mu} q_{kL} + i\overline {u_{kR}}
			\gamma _{\mu}D^{\mu} u_{kR} +i\overline {d_{kR}} \gamma _{\mu}D^{\mu} d_{kR} +\\
			  && + i\overline {l_{kL}} \gamma _{\mu}D^{\mu} l_{kL} + i\overline {e_{kR}}
			\gamma _{\mu}D^{\mu} e_{kR} +i\overline {\nu _{kR}} \gamma _{\mu}D^{\mu} \nu
			_{kR} \Big)  \, ,
\end{eqnarray}
 where the fields' charges are the usual SM and $B-L$ ones (in particular, $B-L = 1/3$ for quarks and $-1$ for leptons {with no distinction between generations, hence ensuring universality)}.
  The  $B-L$ charge assignments of the fields
  as well as the introduction of new
  fermionic  RH heavy neutrinos ($\nu_R$'s) and a
  scalar Higgs field ($\chi$, charged $+2$ under $B-L$)  
  are designed to eliminate the triangular $B-L$  gauge anomalies and to ensure the gauge invariance of the theory, respectively.
  Therefore, the $B-L$  gauge extension of the SM gauge group
  broken at the Electro-Weak (EW) scale does necessarily require
  at least one new scalar field and three new fermionic fields which are
  charged with respect to the $B-L$ group.

Finally, the Yukawa interactions are:
\begin{eqnarray}\nonumber
\mathscr{L}_Y &=& -y^d_{jk}\overline {q_{jL}} d_{kR}H 
                 -y^u_{jk}\overline {q_{jL}} u_{kR}\widetilde H 
		 -y^e_{jk}\overline {l_{jL}} e_{kR}H \\ \label{L_Yukawa}
	      & & -y^{\nu}_{jk}\overline {l_{jL}} \nu _{kR}\widetilde H 
	         -y^M_{jk}\overline {(\nu _R)^c_j} \nu _{kR}\chi +  {\rm 
h.c.}  \, ,
\end{eqnarray}
{where $\widetilde H=i\sigma^2 H^*$ and  $i,j,k$ take the values $1$ to $3$},
where the last term is the Majorana contribution and the others the usual Dirac ones.

Neutrino mass eigenstates, obtained after applying the see-saw mechanism, will be called $\nu_l$ 
(with $l$ standing for light) and $\nu_h$
(with $h$ standing for heavy), where the first ones are the SM-like ones.

\subsection{Realistic models}
The generic model that has been previously introduced spans over a continuous set of minimal $U(1)$ extensions of the SM, that can be labelled by the properties of the charge assignments to the particle content. Notice that such models are, by construction, all and only those that are not anomalous with the SM fermion content
 augmented by one RH neutrino per generation. Therefore, many $E_6$-inspired $U(1)$ scenarios, such as $U(1)_\psi$ or $U(1)_\eta$, are not included in our generic model.

Free parameters in our parameterisation are those appearing in the covariant derivative of eq.~(\ref{cov_der}). We recall the Abelian part only:
\begin{equation}\nonumber
D_{\mu}\equiv \dots +ig_1YB_{\mu} +i(\widetilde{g}Y + g_1'Y_{B-L})B'_{\mu}\, .
\end{equation}
This form of the covariant derivative can be re-written defining an effective coupling $Y^E$ and an effective charge $g_E$:
\begin{equation}\label{eff_par}
g_E Y^E = \, \widetilde{g}Y + g_1'Y_{B-L}.
\end{equation}

As any other parameter in the Lagrangian, $\widetilde{g}$ and $g_1'$ are running parameters \cite{Aguila-Coughlan,CPW}, therefore their values ought to be defined at some scale. A discrete set of popular $Z'$ models (see, e.g., Refs.~\cite{Carena:2004xs,Appelquist:2002mw}) can be recovered by a suitable definition of both $\widetilde{g}$ and $g_1'$.

We will focus our numerical analysis on the scalar sector of the ``pure'' $B-L$ model, that is defined by the condition $\widetilde{g}(Q_{EW}) = 0$, i.e., we nullify it at the EW scale. This implies no mixing at the tree-level between the  $B-L$ $Z'$ and SM $Z$ gauge bosons. Other benchmark models of our general parameterisation are for example the Sequential SM (SSM), defined by $Y^E = Y$ (that in our notation corresponds to the condition $g'_1=0$ at the EW scale) and the $U(1)_R$ model, for which RH fermion charges vanish (that is recovered here by the condition $\widetilde{g}=-2g'_1$ at the EW scale).

It is important to note that none of the models described so far is orthogonal to the $U(1)_Y$ of the SM, therefore the RGE running of the fundamental parameters, $\widetilde{g}$ and $g_1'$, will modify the relations above. The only orthogonal $U(1)$ extension of the SM is the ``$SO(10)$-inspired'' $U(1)_\chi$ model, that in our notation reads $\widetilde{g}=-\frac{4}{5}g_1'$. Although the $\widetilde{g}$ and $g_1'$ couplings run with a different behaviour, the EW relation $\widetilde{g}/g_1' = -4/5$ is preserved (at one-loop) at any scale.

Nonetheless, as indeed true for the SM, the gauge sector affects marginally the scalar sector in its running, so the analysis we are going to show is effectively independent of the specific charge assignation. However, we might expect differences regarding the interplay between the gauge sector and the neutrino evolution, that impinge on the vacuum stability studies of the scalar sector as the top quark does for the SM Higgs sector. We will report separately on the study of the differences in the RGE study of the specific benchmark models in our generic parameterisation.

\section{Computational details}\label{sect:comp_det}
The complete set of RGEs for the generic model are derived for the parameters in the Lagrangian and are collected in appendix~\ref{RGs}. 
For their numerical study, we put boundary conditions at the EW scale on the physical observables: $m_{h_1},\, m_{h_2},\, \alpha,\, v, M_{Z'}, g'_1, \widetilde{g}, m^{1,2,3}_{\nu_h}$, that we trade for $m,\, \mu,\, \lambda _1,\, \lambda _2,\, \lambda _3,\, x, y^M_{1,2,3}$ using, for the relevant parameters therein, eq.~(\ref{inversion}). Where stated in the text, we impose boundary conditions on some parameters of the Lagrangian rather than on the physical observables. This is done for consistency of those studies.

For the pure $B-L$ model, object of the numerical analysis in this work, the definition $\widetilde{g}=0$ holds, and as a consequence, we also have that the $B-L$ breaking VEV $x$ can be easily related to the new $Z'$ boson mass by $\displaystyle x=\frac{M_{Z'}}{2g'_1}$, where we fixed $g'_1=0.1$. Regarding the neutrinos, for simplicity we consider them degenerate and we fix their masses to $m^{1,2,3}_{\nu_h} \equiv m_{\nu_h} = 200$ GeV (whenever not specified otherwise), a value that can lead to some interesting phenomenology \cite{bbms}. The free parameters in our study are then $m_{h_1}$, $m_{h_2}$, $\alpha$ and $x$. The general philosophy is to fix in turn some of the free parameters and scan over the other ones, individuating the allowed regions fulfilling the following set of conditions. 

We first define a parameter to be ``perturbative'' for values less than unity. This is a conservative definition, as we could relax it by an order of magnitude and still get values of the parameters for which the perturbative series will converge\footnote{Notice that, in analogy with QED, the parameters upon which the perturbative expansion is performed are usually of the form $\sqrt{\alpha} = g/ \sqrt{4\pi}$, rather then being $g$ itself.}.
RGE evolution can then constrain the parameter space of the scalar sector in two complementary ways. From one side, the couplings must be perturbative. This condition reads:
\begin{equation}\label{cond_1}
0 < \lambda _{1,2,3}(Q') < 1 \qquad \forall \; Q' \leq Q\, ,
\end{equation}
and it is usually referred to as the ``triviality'' condition.
On the other side, the vacuum of the theory must be well-defined at any scale, that is, to guarantee the validity of eqs.~(\ref{inf_limitated}) and (\ref{positivity}) at any scale $Q'\leq Q$:
\begin{equation}\label{cond_2}
0 < \lambda _{1,2,3}(Q') \qquad \mbox{and} \qquad
4\lambda _1(Q')\lambda _2(Q')-\lambda _{3}^2(Q') > 0 \qquad \forall \; Q' \leq Q\, .
\end{equation}
Eq.~(\ref{cond_2}) is usually referred to as the ``vacuum stability'' condition. In contrast to the SM, in which it is sufficient the Higgs self-coupling $\lambda$ be positive, in the case of this model the vacuum stability condition (and especially the second part of eq.~(\ref{cond_2})) can be violated even for positive $\lambda _{1,2,3}$.

One should notice that our conventional choice $m_{h_1} < m_{h_2}$, as noted previously, let us consider $\alpha$ and $-\alpha$ as two independent solutions, although the theory is manifestly invariant under the symmetry $\alpha\rightarrow -\alpha$. These two solutions are complementary, meaning that the region excluded by the choice $m_{h_1} < m_{h_2}$ at a certain value of the angle $\alpha$ is precisely the allowed one for the complementary angle $\pi /2 - \alpha$. The special case $\alpha=\pi /4$ is symmetric, and corresponds to maximal mixing between the scalars. $\alpha=0$ corresponds to a SM scalar sector totally decoupled from the extended one, and $h_1$ is the usual SM Higgs boson. $\alpha=\pi /2$ is the specular case, in which $h_2$ plays the role of the SM Higgs boson.

Notice also that, again in contrast to the SM in which the gauge couplings have a marginal effect, in our case the RH neutrinos play for the extra scalar singlet the role of the top quark for the SM Higgs in the vacuum stability condition\footnote{Also notice that we have three RH neutrinos, as we have three colours for the top quark. However, they are Majorana particles rather than Dirac ones, so they carry half (independent) degrees of freedom than the top quark.}. Their RGE are then controlled by the Yukawa coupling with a negative contribution coming from $g'_1$ (see eq.~(\ref{RGE_nu_r_maj})). Therefore, in some regions of the parameter space, the impact of the gauge sector is not marginal and can effectively stabilise the otherwise divergent evolution of the Majorana Yukawa couplings for the RH neutrinos. We will report on the effect of RH neutrinos in our analysis in section~\ref{sect:neutrino_eff}.

A final remark is in order about eq.~(\ref{RGE_lamda3}), the evolution of $\lambda _3$,  the mixing parameter of the scalar potential (see eq.~(\ref{BL-potential})). This RGE is almost proportional to $\lambda _3$ itself, so a vanishing boundary condition is almost stable\footnote{From the last line of eq.~(\ref{inversion}), setting $\lambda _3 =0$ corresponds to $\alpha =0$, but not vice versa.}. Non-proportional terms arise from the new gauge couplings ($\widetilde{g}$ and $g'_1$), i.e., deviations from the vanishing boundary conditions are of the order of the gauge coupling, hence quite small. They are particularly negligible in the pure $B-L$ model, as also $\widetilde{g}$ has a vanishing boundary condition, with a weak departure from it due to the mixing in the gauge coupling sector \cite{B-L}. Nonetheless, other benchmark models in our general parameterisation could show different behaviours.

\section{Results}\label{sect:results}
We present here our results for the pure $B-L$ model, the chosen benchmark of our general parameterisation. We will first present a brief analysis of the gauge sector, followed by a quick review of the present experimental constraints on the Higgs boson masses. Finally, we will fully describe the scalar sector analysis, argument of this paper\footnote{Notice that we study the gauge sector of the model (and, in particular, the Abelian part of it) independently of any other sector as the
corresponding RGEs fully decouple.}.

\subsection{Gauge sector}
Before starting the analysis of the scalar sector, we can briefly look at the gauge sector, where the RGE evolution gives us indications for the validity of the model concerning the gauge couplings. In particular, their evolution must stay perturbative up to some particular scale. In the $B-L$ model, the conditions that the free parameters in the gauge sector must fulfil are:
\begin{equation}\label{cond_g}
g'_1 (Q') < 1 \hspace{0.5cm} \forall\; Q'\leq Q \qquad \mbox{and} \qquad \widetilde{g}(Q_{EW}) = 0\, ,
\end{equation}
where the second condition in eq.~(\ref{cond_g}) defines the pure $B-L$ model.

\begin{figure}[!h]
  \centering
  \includegraphics[angle=0,width=0.48\textwidth ]{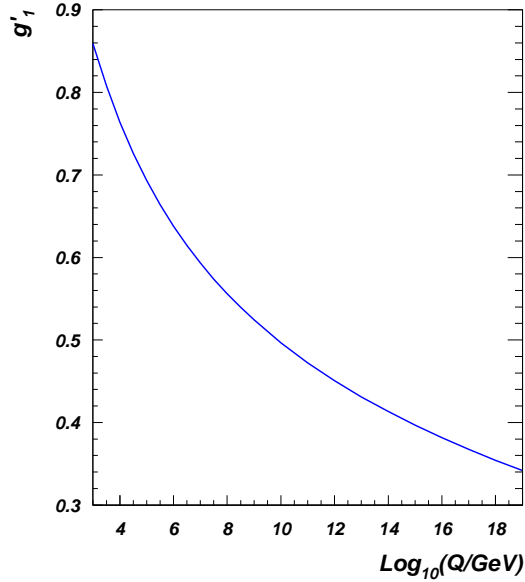}
  \vspace*{-0.5cm}
  \caption{\it Maximum allowed values by eq.~(\ref{cond_g}) for $g'_1(Q_{EW})$  in the $B-L$ model 
as a function of the scale $Q$.   
\label{g1p_vs_Q}}
\end{figure}

Varying the scale $Q$, the maximum scale up to which we want the model to be well-defined, we get an 
upper bound on $g'_1(Q_{EW})$ as a function of $Q$, as shown in figure~\ref{g1p_vs_Q}. 
Typical results are summarised in table~\ref{g1p-up_bound}.

\vspace*{0.5cm}
\begin{table}[h]
\begin{center}
\begin{tabular}{|c|c|c|c|c|c|c|}
\hline
$Log_{10} (Q/ \mbox{GeV}) $ & 3 & 5 & 7 & 10 & 15 & 19  \\  \hline
$g_1'(Q_{EW})$  & 0.860 & 0.693 & 0.593 & 0.497 & 0.397 & 0.342  \\ 
\hline
\end{tabular}
\end{center}
\vskip -0.5cm
\caption{\it Maximum allowed values by eq.~(\ref{cond_g}) for $g'_1(Q_{EW})$ in the $B-L$ model 
for selected values of the scale $Q$.
\label{g1p-up_bound}}
\end{table}

\subsection{Experimental limit}\label{sect:exp-lim}

\begin{figure}[!h]
  \centering
  \includegraphics[angle=0,width=0.48\textwidth ]{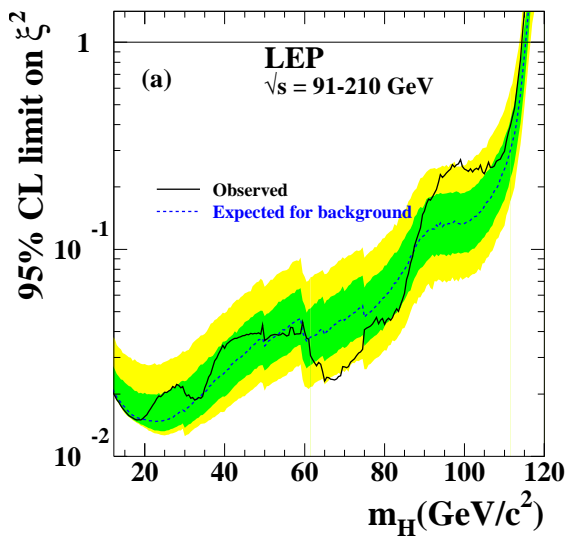}
  \vspace*{-0.5cm}
  \caption{\it The $95\%$ C.L. upper bound on $\xi=g_{HZZ}/g^{SM}_{HZZ}$ \cite{Barate:2003sz}. In the $B-L$ model, $\xi=\cos{\alpha}(\sin{\alpha})$ for $H=h_1(h_2)$.  \label{LEP_lim}}
\end{figure}

Past and current experiments have set limits on the scalar sector parameters in the SM as well as in various 
extensions of it, see for example Ref.~\cite{Barate:2003sz} for LEP and Ref.~\cite{Tevatron_bounds} for Tevatron. 
For the model discussed here, the relevant analysis is summarised in figure~\ref{LEP_lim}, in which a generic overall factor $\xi$ has been introduced. 
Such parameter is defined as the coupling(s) to the $Z$ boson of the Higgs particle(s) in the considered extension 
normalised to the SM:
\begin{equation}
\xi\equiv \frac{g_{HZZ}}{g^{SM}_{HZZ}}\, ,
\end{equation}
hence it parametrises the deviations of the new model with respect to the SM.

In the minimal $U(1)$ extension of the SM, argument of this paper, two scalar eigenstates exist: 
the one coming from the Higgs singlet, required to break the extra $U(1)_{B-L}$ gauge factor 
(and therefore giving the $Z'$ gauge boson a mass), and the one coming from the Higgs doublet, 
required to break the SM gauge symmetry to give masses to the $W$ and $Z$ bosons. 
With reference to eq.~(\ref{scalari_autostati_massa}), we called $h_1$ the lightest of such eigenstates, 
that couples to the $Z$ boson proportionally to $\cos{\alpha}$, and with $h_2$ we referred to the heaviest 
scalar, that couples to the $Z$ boson proportionally to $\sin{\alpha}$. Hence, the LEP lower bounds on 
the scalar masses of the $U(1)_{B-L}$ extension here considered are read straightforwardly from figure~\ref{LEP_lim} 
by considering:
\begin{equation}
\left\{ \begin{array}{cc} 
	\xi=\cos{\alpha} &\qquad \mbox{for } H = h_1\, ,\\
	\xi=\sin{\alpha} &\qquad \mbox{for } H = h_2\, ,
	\end{array} \right. 
\end{equation}
i.e., the limit for $h_1$($h_2$) are extracted by considering $\xi$ as the cosine(sine) of the mixing angle 
in the scalar sector (see eq.~(\ref{scalari_autostati_massa}) and the following ones).

Figure~\ref{LEP_lim} shows the lower bound on the Higgs mass as a function of $\xi$. The SM Higgs is 
recovered by the condition $\xi =1$. We see that we can have significant deviation from the SM Higgs mass limit, 
$m_h > 114.4$ GeV, only for values of the angle $\alpha > \pi /4$, for the lightest state $h_1$. For example, 
for $\alpha = \pi /3$, the LEP limit on the lightest Higgs state reads as $m_{h_1} > 100$ GeV. That is, in this 
model, a light Higgs with mass smaller than the SM limit can exist only if it is highly mixed, i.e., the light 
Higgs is mostly the singlet state. For the same value of the angle, the limit for $m_{h_2}$ is more stringent 
than the condition $m_{h_2} > m_{h_1}$, in fact for $\alpha = \pi /3$, $m_{h_2} \gtrsim 114$ GeV must be fulfilled.

The LEP experiments are also able to provide a lover bound for the $B-L$ breaking VEV $x$. In fact, the LEP 
bound on the $B-L$ $Z'$ mass \cite{Cacciapaglia:2006pk},
\begin{equation}\label{LEP_bound}
\frac{M_{Z'}}{g'_1} \geq 7\; \rm{TeV}\, 
\end{equation}
can be rewritten as a lower bound for the VEV:
\begin{equation}\label{LEP_bound_vev}
x \geq 3.5\; \rm{TeV}\, ,
\end{equation}
since $M_{Z'}=2xg'_1$ in the pure $B-L$ model.

\subsection{Scalar sector}
Given the simplicity of the scalar sector in the SM, the triviality and vacuum stability conditions can be studied independently and they both constrain the Higgs boson masses, providing an upper bound and a lower bound, respectively. In more complicated models as the one considered here, it might be more convenient to study the overall effect of eqs.~(\ref{cond_1})-(\ref{cond_2}), since there are regions of the parameter space in which the constraints are evaded simultaneously. This is the strategy we decided to follow.

\begin{figure}[!h]
  \subfloat[]{ 
  \label{mh1_mh2_a0}
  \includegraphics[angle=0,width=0.48\textwidth ]{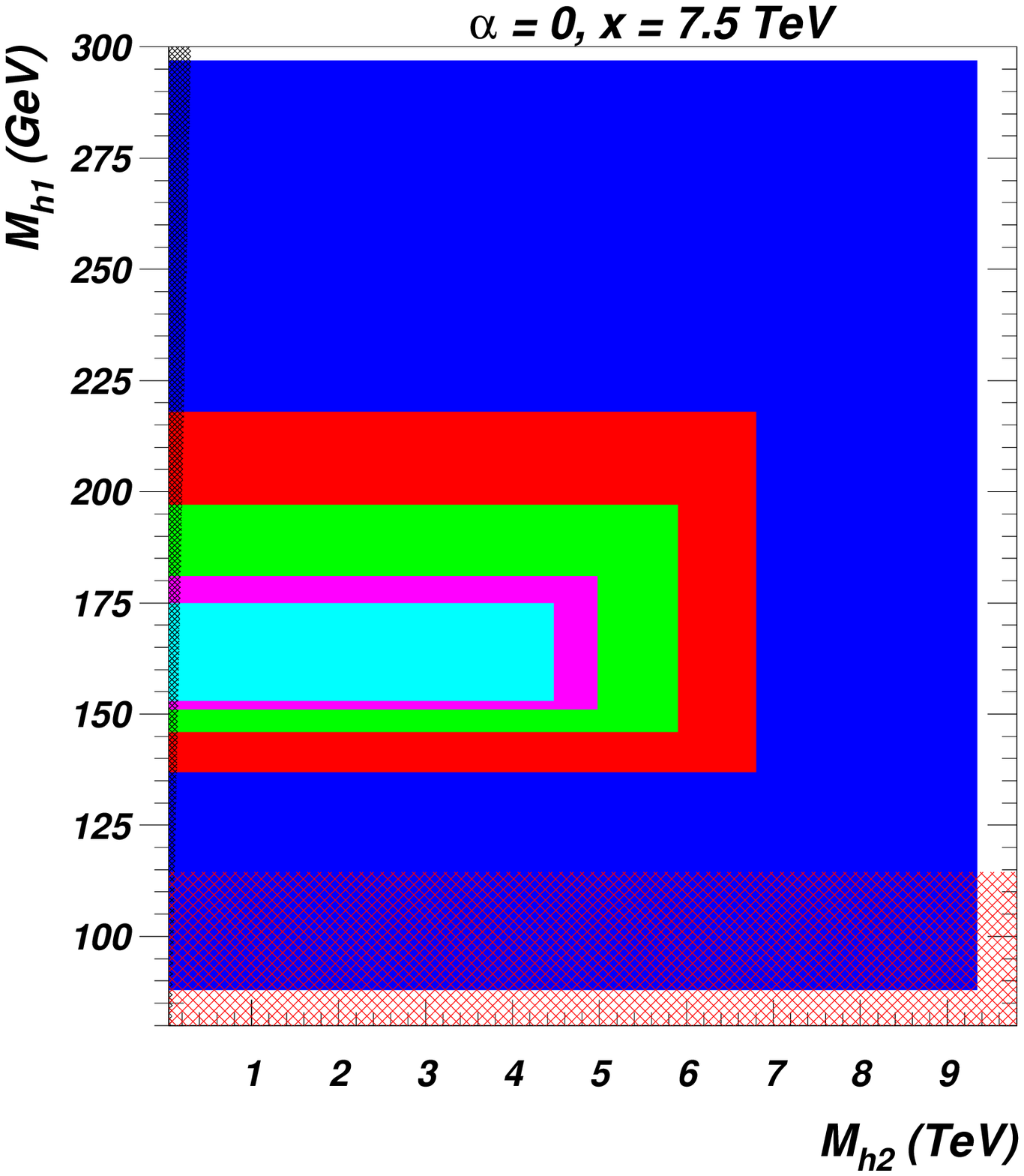}}
  \subfloat[]{
  \label{mh1_mh2_a01}
  \includegraphics[angle=0,width=0.48\textwidth ]{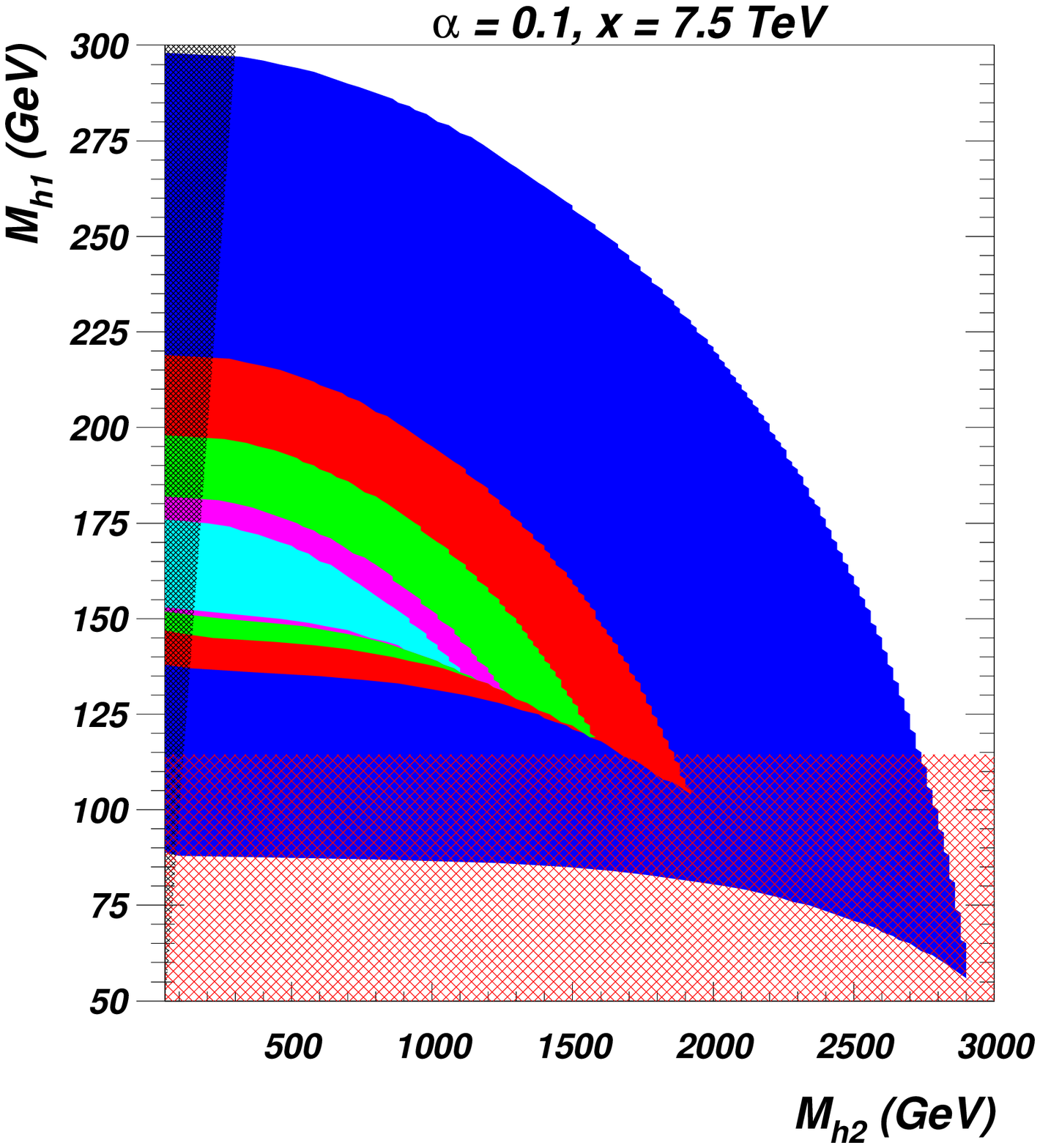}}
  \\
  \subfloat[]{
  \label{mh1_mh2_api4}
  \includegraphics[angle=0,width=0.48\textwidth ]{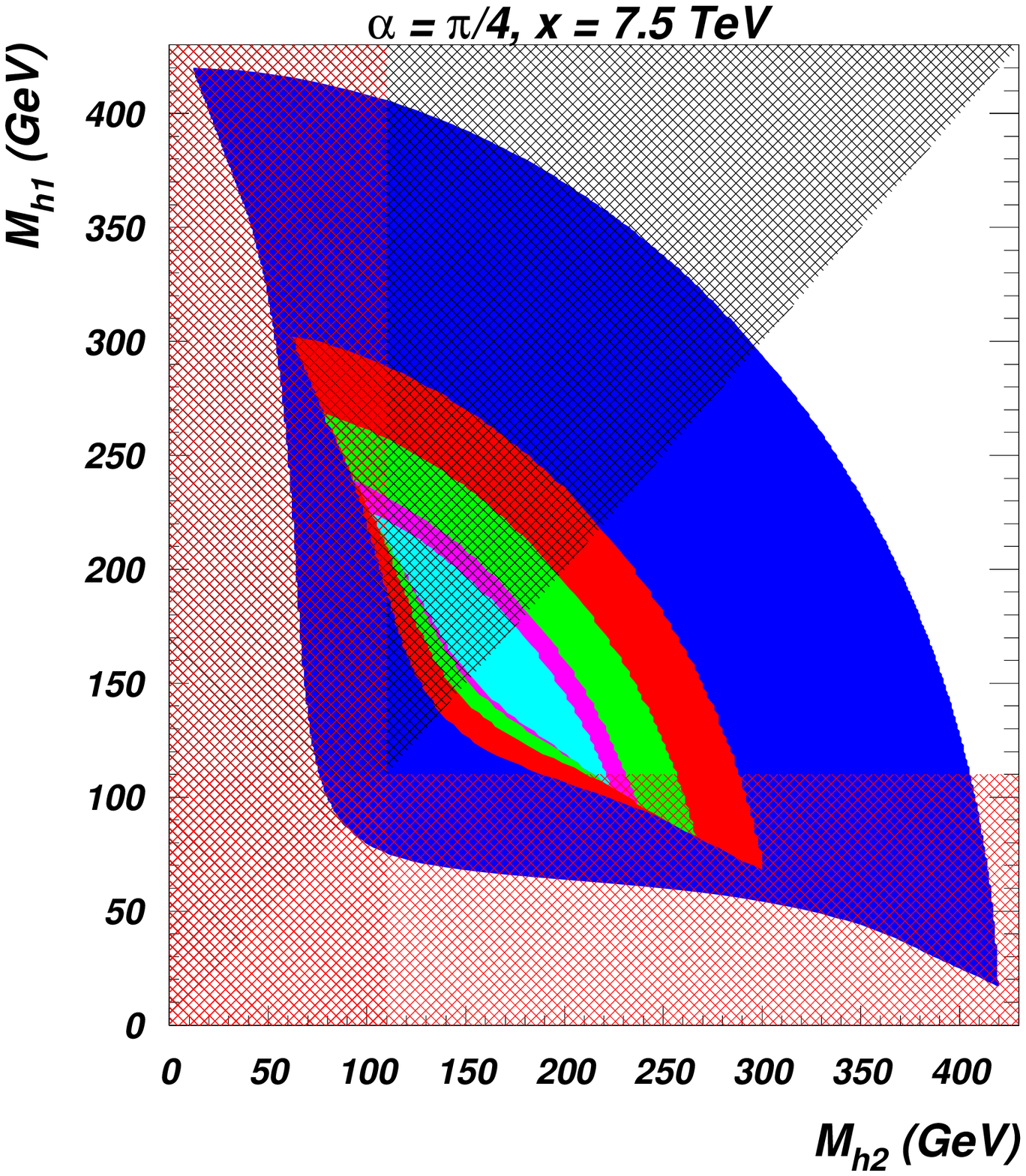}}
  \subfloat[]{
  \label{mh1_mh2_api3}
  \includegraphics[angle=0,width=0.48\textwidth ]{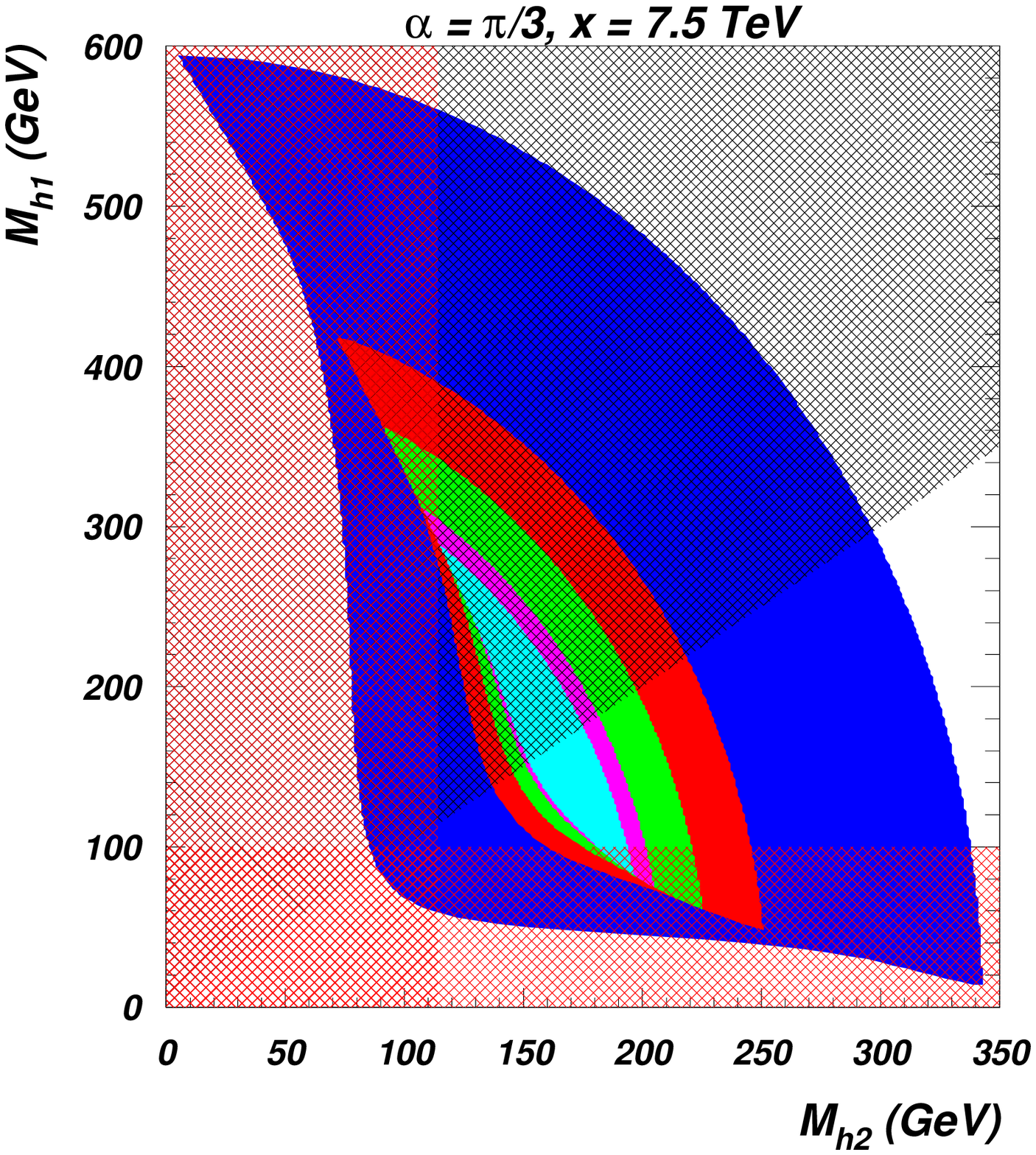}}
  \vspace*{-0.5cm}
  \caption{\it Allowed values in the $m_{h_1}$ vs. $m_{h_2}$ space in the $B-L$ model by eqs.~(\ref{cond_1}) and (\ref{cond_2}), for (\ref{mh1_mh2_a0}) $\alpha =0$, (\ref{mh1_mh2_a01}) $\alpha =0.1$, (\ref{mh1_mh2_api4}) $\alpha =\pi /4$ and (\ref{mh1_mh2_api3}) $\alpha =\pi /3$. Colours refer to different values of $Q/$GeV: blue ($10^{3}$), red ($10^{7}$), green ($10^{10}$), purple ($10^{15}$) and cyan ($10^{19}$). The shaded black region is forbidden by our convention $m_{h_2} > m_{h_1}$, while the shaded red region refers to the values of of the scalar masses forbidden by LEP. Here: $x=7.5$ TeV,
$m_{\nu_h}=200$ GeV.  \label{mh1_mh2}}
\end{figure}

Figure~\ref{mh1_mh2} shows the allowed region in the parameter space $m_{h_1}$-$m_{h_2}$ for increasing values of the mixing angle $\alpha$, for fixed VEV $x=7.5$ TeV and heavy neutrino masses $m_{\nu_h}=200$ GeV, corresponding to Yukawa couplings whose effect on the RGE running can be considered negligible.
For $\alpha=0$, the allowed values for $m_{h_1}$ are the SM ones and the extended scalar sector is completely decoupled. The allowed space is therefore the simple direct product of the two, as we can see in figure~\ref{mh1_mh2_a0}. When there is no mixing, the bounds we get for the new heavy scalar are quite loose, allowing a several TeV range for $m_{h_2}$, depending on the scale of validity of the theory. We observe no significant lower bounds (i.e., $m_{h_2}>0.5$ GeV), as the RH Majorana neutrino Yukawa couplings are negligible.

As we increase the value for the angle, the allowed space deforms towards smaller values of $m_{h_1}$. If for very small scales $Q$ of validity of the theory such masses have already been excluded by LEP, for big enough values of $Q$, at a small angle as $\alpha=0.1$, the presence of a heavier boson allows the model to survive up to higher scales for smaller $h_1$ masses if compared to the SM (in which just $h_1$ would exist). Correspondingly, the constraints on $m_{h_2}$ become tighter. Moving to bigger values of the angle, the mixing between $h_1$ and $h_2$ grows up to its maximum, at $\alpha = \pi /4$, where $h_1$ and $h_2$ both contain an equal amount of doublet and singlet scalars. The situation is therefore perfectly symmetric, as one can see from figure~\ref{mh1_mh2_api4}. Finally, in figure~\ref{mh1_mh2_api3}, we see that the bounds on $m_{h_2}$ are getting tighter, approaching the SM ones, and those for $m_{h_1}$ are relaxing. That is, for values of the angle $\pi/4 < \alpha < \pi /2$, the situation is qualitatively not changed, but now $h_2$ is the SM-like Higgs boson. Visually, one can get the allowed regions at a given angle $\pi/2 - \alpha$ by simply taking the transposed about the $m_{h_1}=m_{h_2}$ line of the plot for the given angle $\alpha$.

\begin{figure}[!h]
  \subfloat[]{ 
  \label{mh2_vs_Q_a0}
  \includegraphics[angle=0,width=0.48\textwidth ]{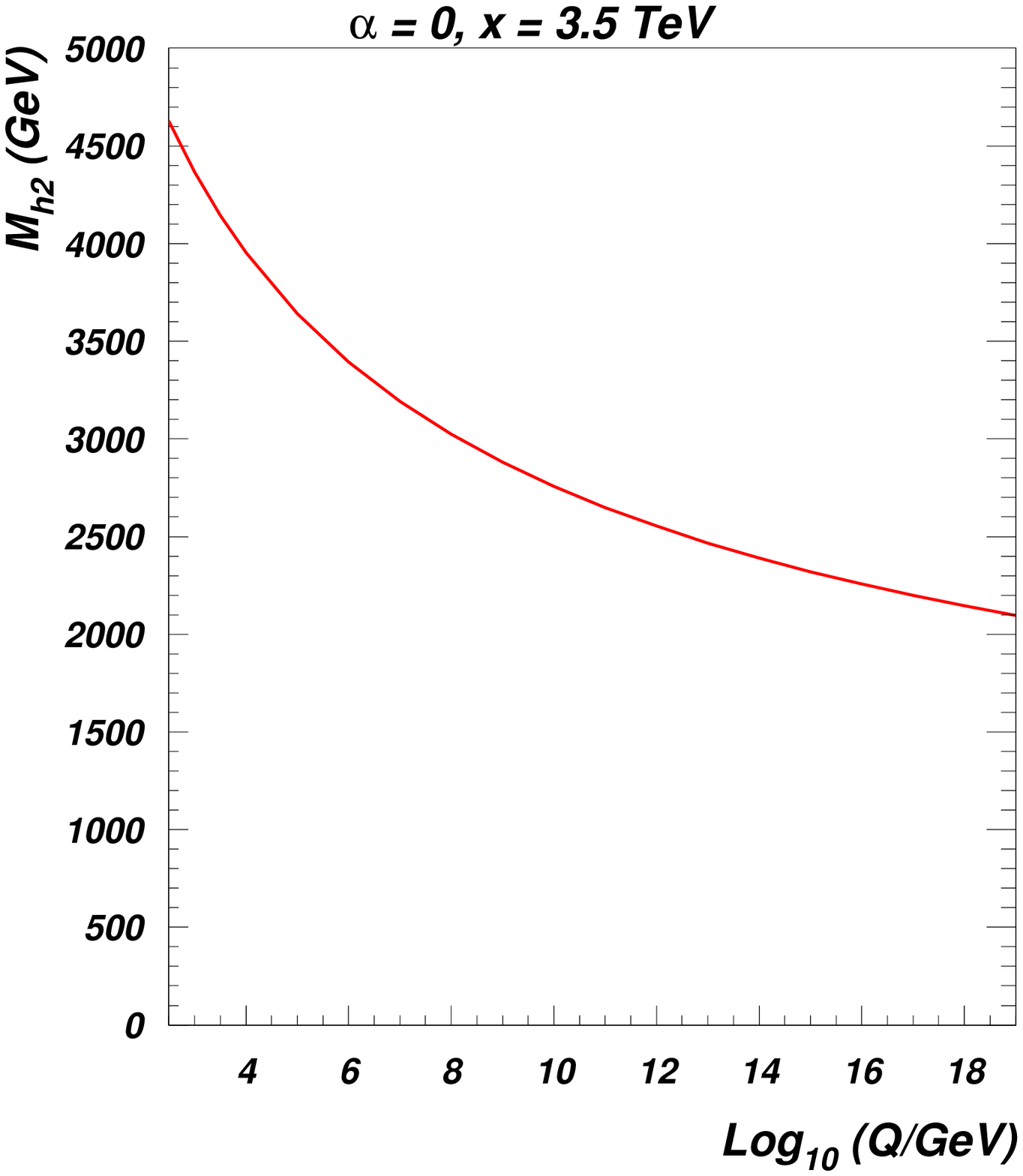}}
  \subfloat[]{
  \label{mh2_vs_Q_a01}
  \includegraphics[angle=0,width=0.48\textwidth ]{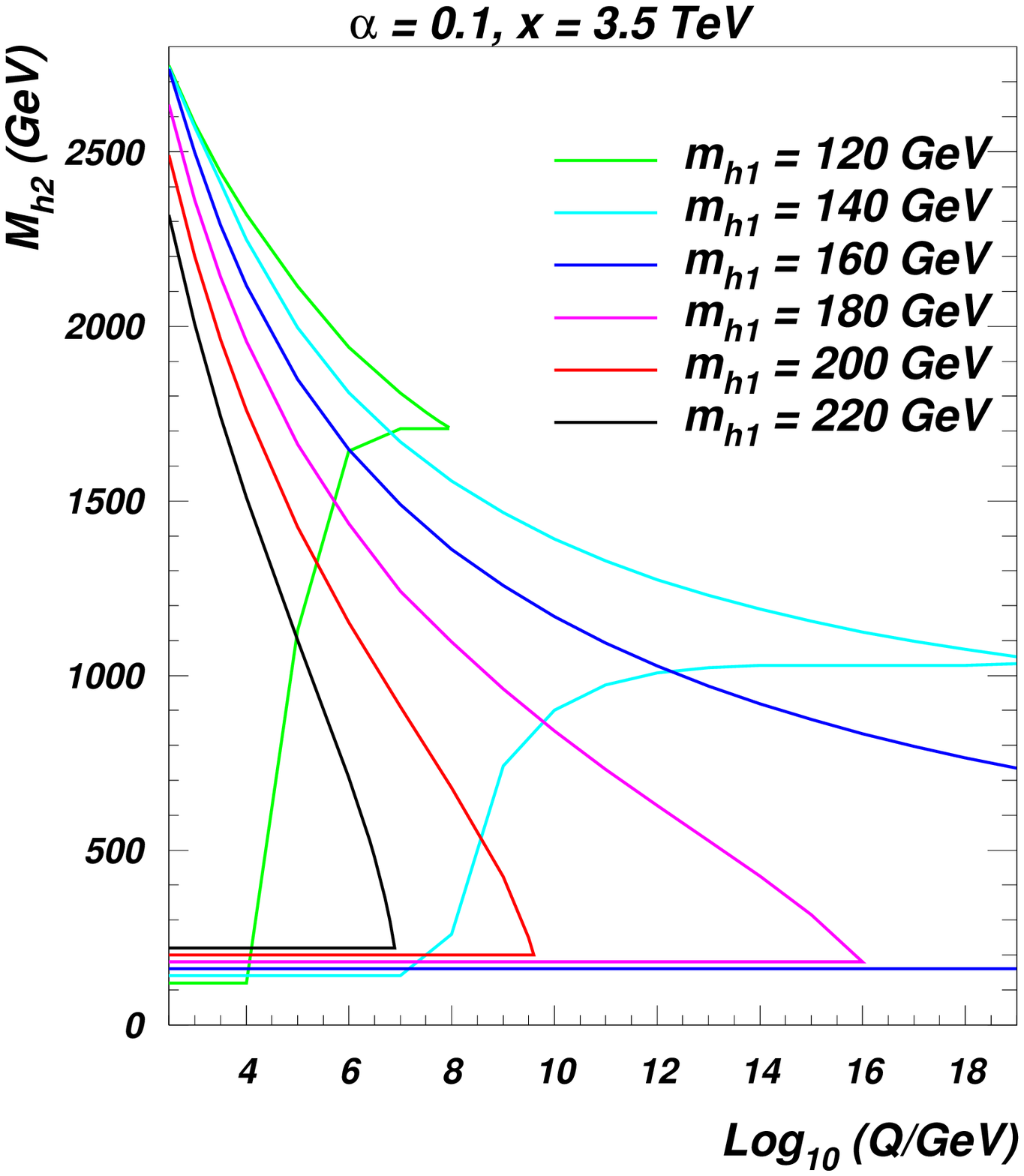}}
  \\
  \subfloat[]{ 
  \label{mh2_vs_Q_pi8}
  \includegraphics[angle=0,width=0.48\textwidth ]{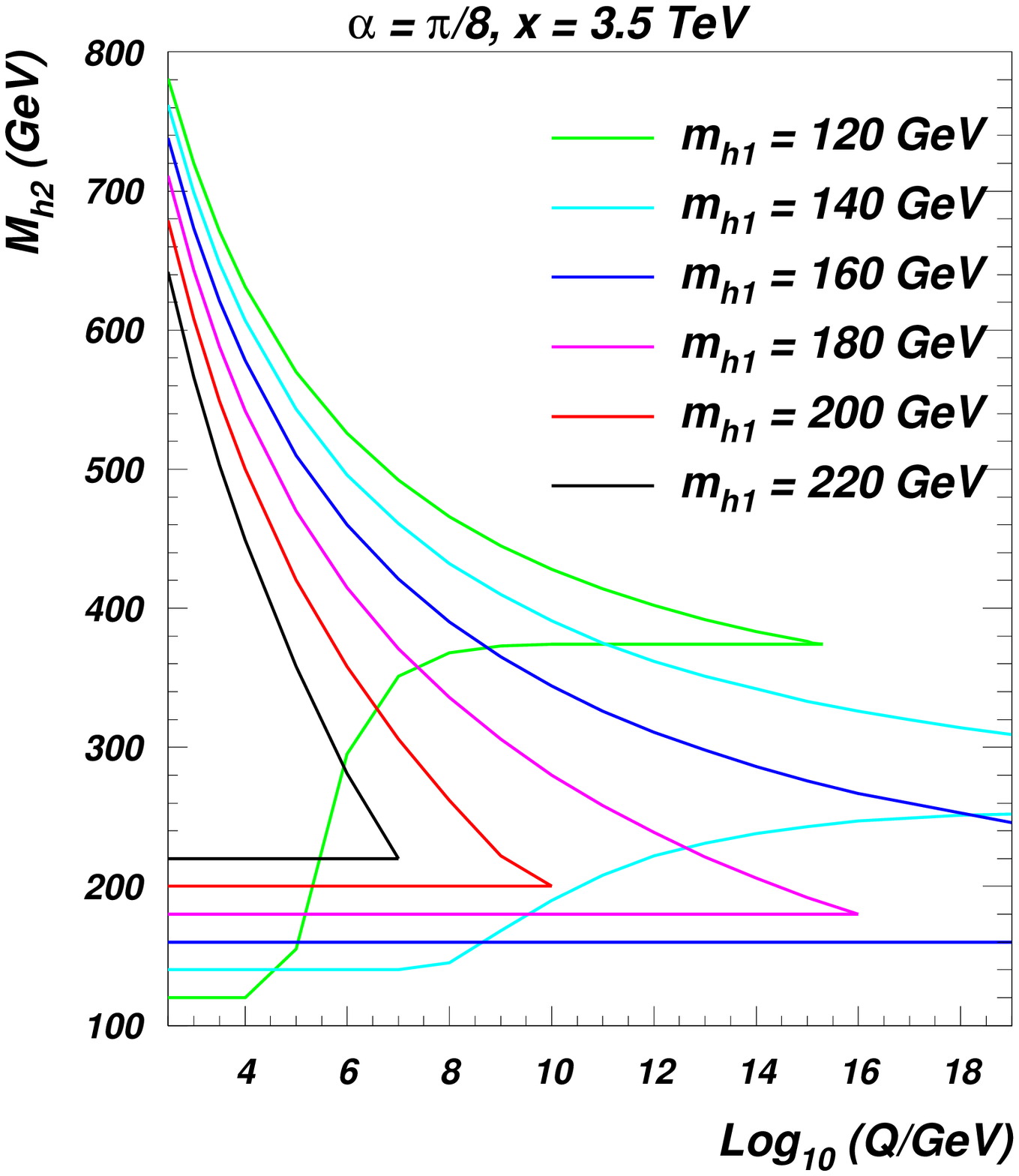}}
  \subfloat[]{
  \label{mh2_vs_Q_api4}
  \includegraphics[angle=0,width=0.48\textwidth ]{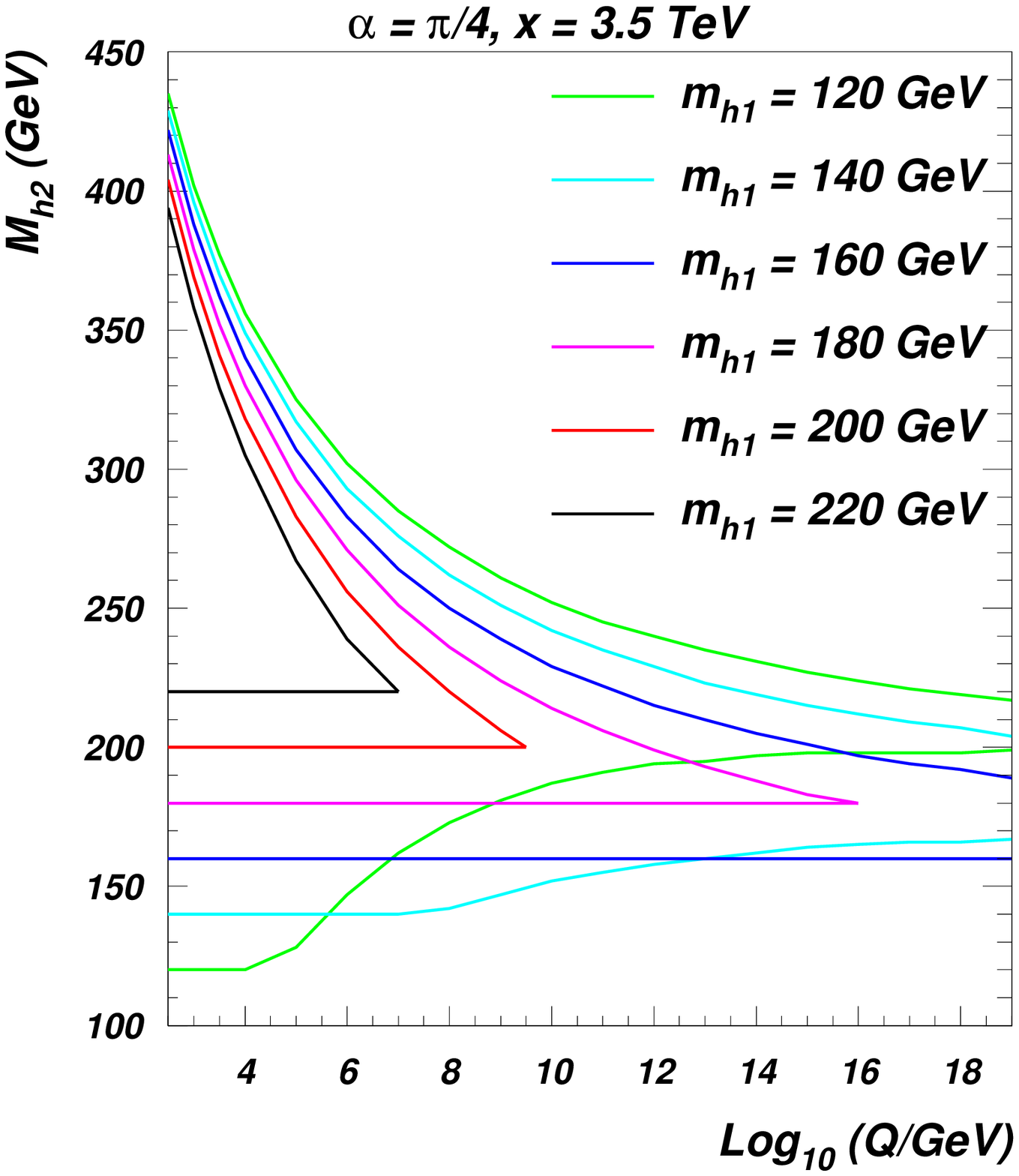}}
    \vspace*{-0.5cm}
  \caption{\it Allowed values (that are those between the same colour lines) for $m_{h_2}$ as a function of the scale $Q$ in the $B-L$ model by eqs.~(\ref{cond_1}) and (\ref{cond_2}), for several values of $m_{h_1}$ and (\ref{mh2_vs_Q_a0}) $\alpha =0$, (\ref{mh2_vs_Q_a01}) $\alpha =0.1$, (\ref{mh2_vs_Q_pi8}) $\alpha =\pi /8$ and (\ref{mh2_vs_Q_api4}) $\alpha = \pi /4$. Also, $x=3.5$ TeV and $m_{\nu_h}=200$ GeV. Only the allowed values by our convention $m_{h_2} > m_{h_1}$ are shown.  \label{mh2_Q}}
\end{figure}

Per each value of the angle, we can then fix the lighter Higgs mass $m_{h_1}$ to some benchmark values (allowed by LEP for the SM Higgs) and plot the allowed mass for the heavier Higgs as a function of the scale $Q$. This is done in figure~\ref{mh2_Q}, where the allowed masses are those contained between the same colour lines. Notice that here the VEV $x$ is fixed to a different value, $x=3.5$ TeV. The effects of changing the VEV $x$ will be described in section~\ref{sect:VEV_eff}.

As previously noticed, the allowed range in $m_{h_2}$ gets smaller as we increase the angle. Apart from the case $\alpha =0$ where there is no dependency at all from $m_{h_1}$, there is a strong effect from $m_{h_1}$ on the bounds on $m_{h_2}$. Not all the allowed regions at a fixed $h_1$ mass are contained in the region for a smaller $m_{h_1}$. This is true only for $m_{h_1} > 160$ GeV. For smaller $m_{h_1}$'s, the distortion in the allowed region constraints tightly $m_{h_2}$ for the survival of the model to big scales $Q$. This is because such distortion is just towards smaller $h_1$ masses, see figure~\ref{mh1_mh2}.

\begin{figure}[!h]
  \subfloat[]{ 
  \label{mh2_a_mh1-100}
  \includegraphics[angle=0,width=0.48\textwidth ]{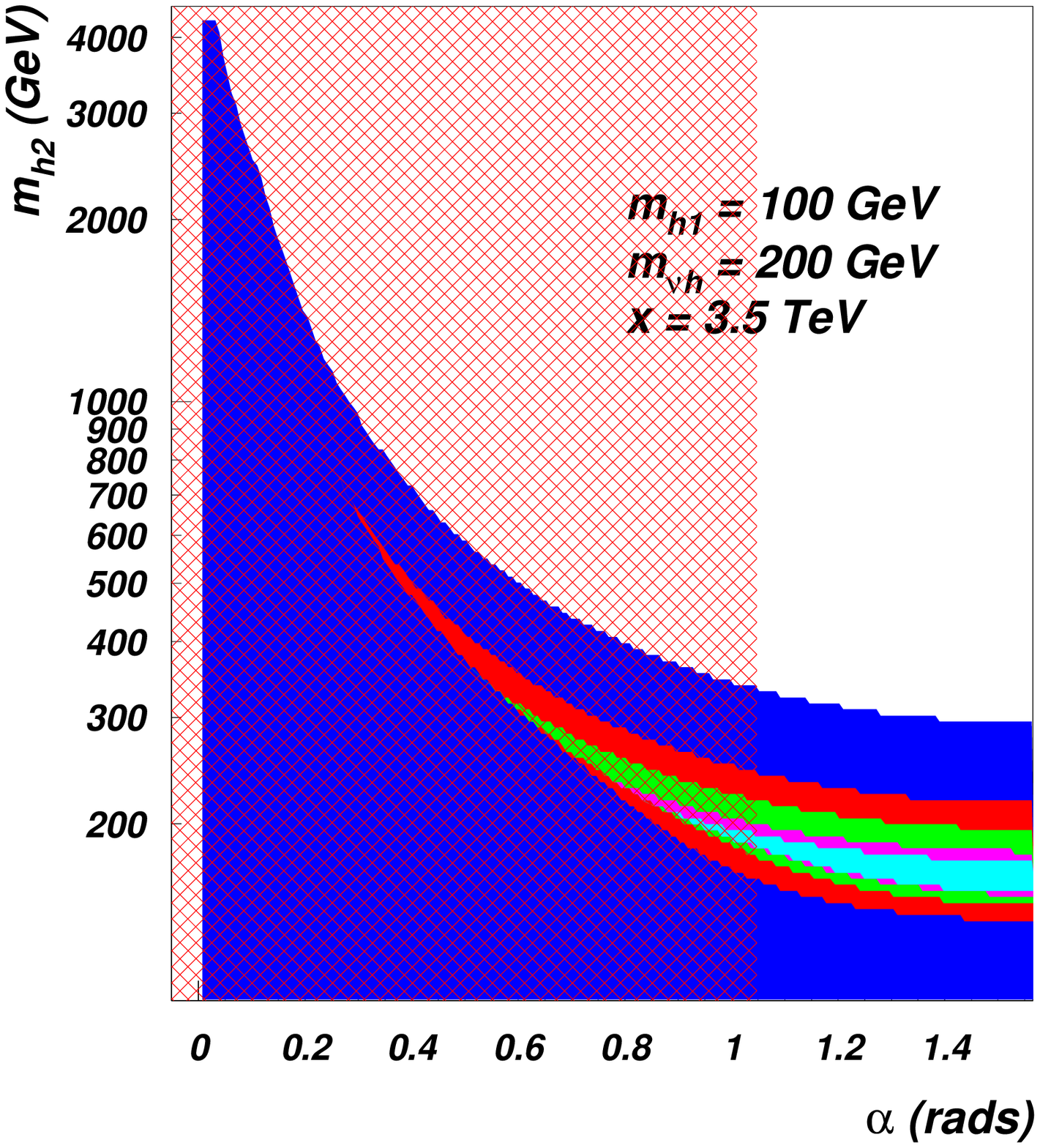}}
  \subfloat[]{
  \label{mh2_a_mh1-120}
  \includegraphics[angle=0,width=0.48\textwidth ]{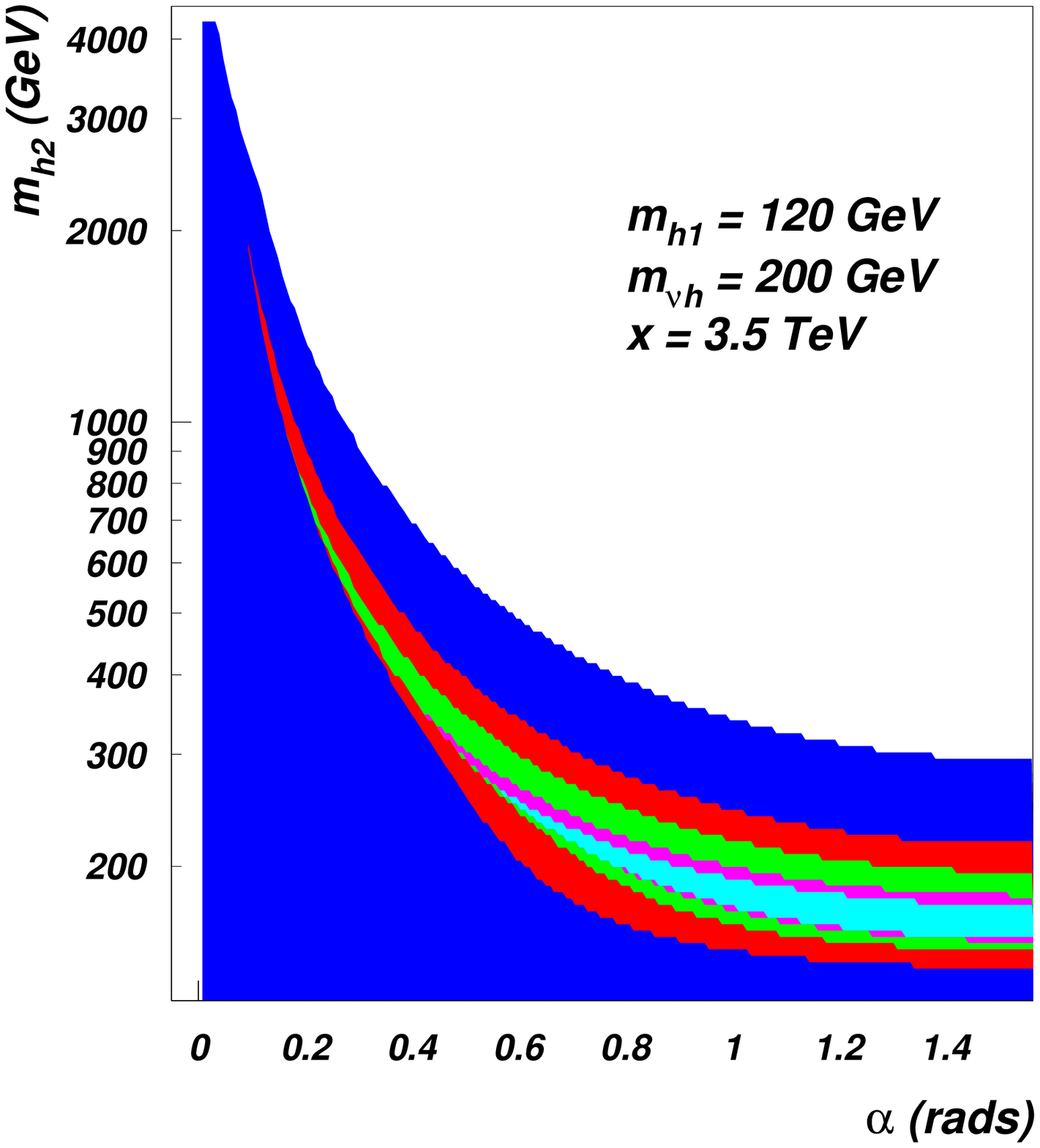}}
\\
  \subfloat[]{
  \label{mh2_a_mh1-160}
  \includegraphics[angle=0,width=0.48\textwidth ]{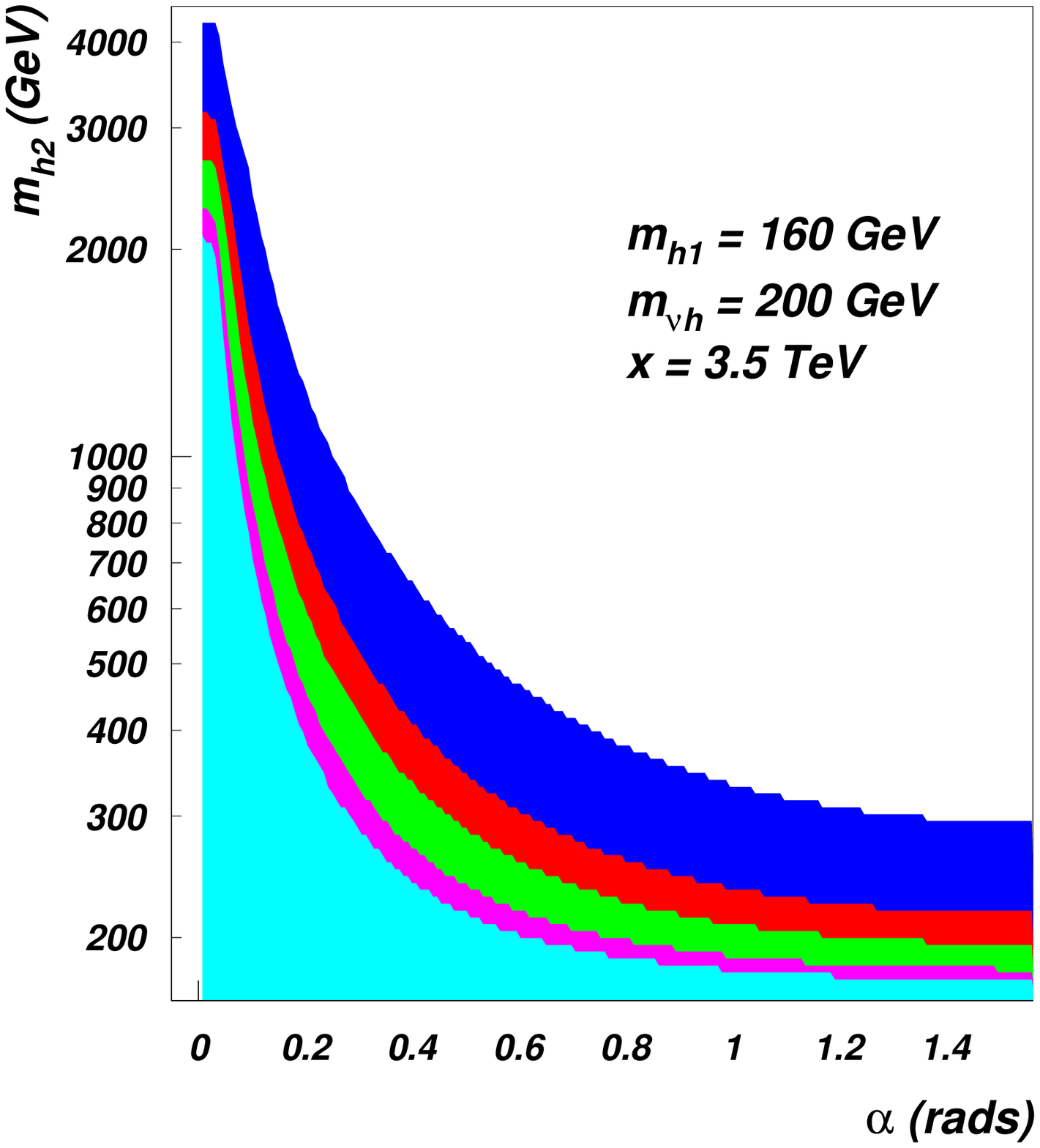}}
  \subfloat[]{
  \label{mh2_a_mh1-180}
  \includegraphics[angle=0,width=0.48\textwidth ]{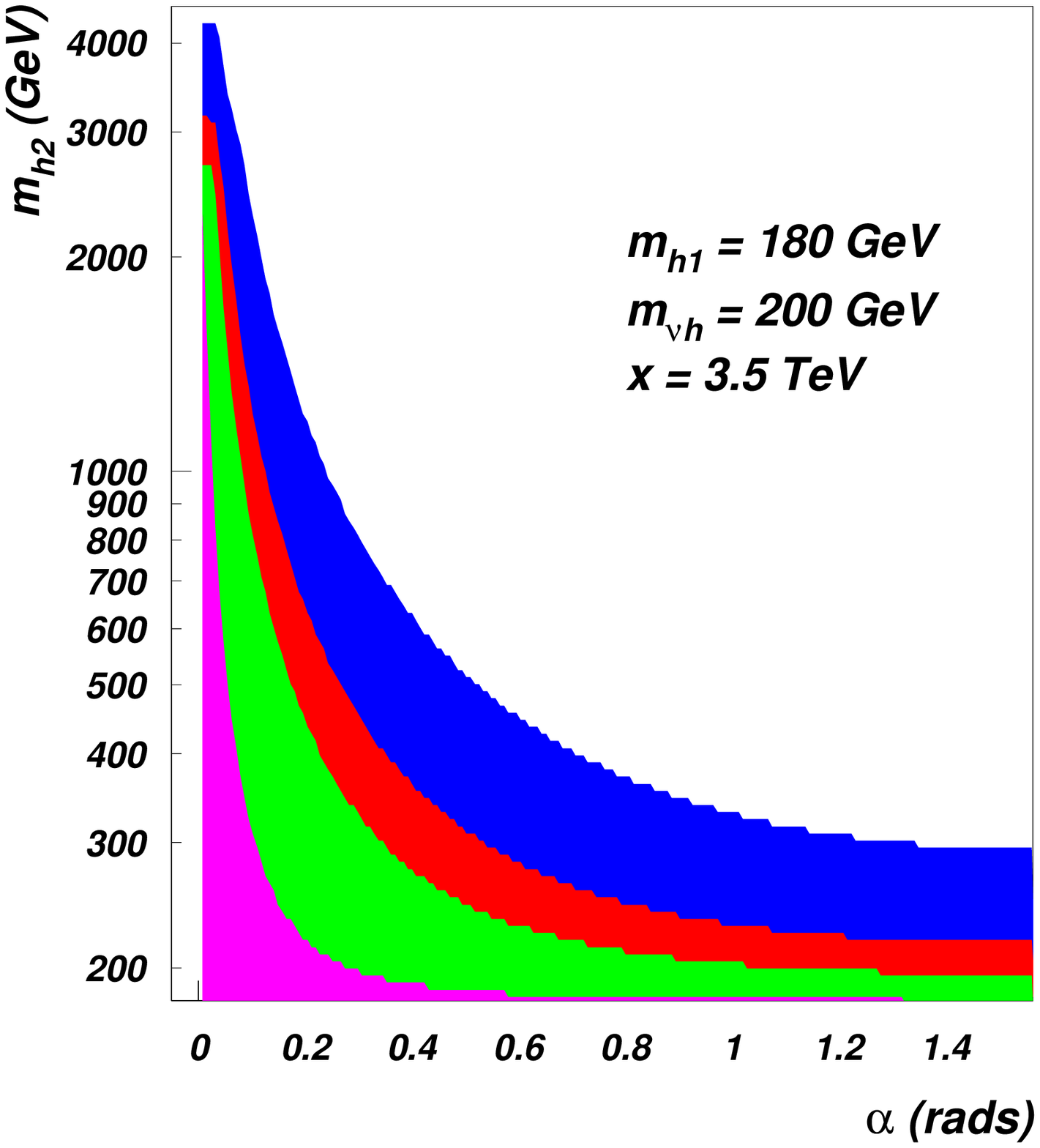}}
  \vspace*{-0.5cm}
  \caption{\it Allowed values in the $m_{h_2}$ vs. $\alpha$ space in the $B-L$ model by eqs.~(\ref{cond_1}) and (\ref{cond_2}), for (\ref{mh2_a_mh1-100}) $m_{h_1}=100$ GeV, (\ref{mh2_a_mh1-100}) $m_{h_1}=120$ GeV, (\ref{mh2_a_mh1-100}) $m_{h_1}=160$ GeV and (\ref{mh2_a_mh1-100}) $m_{h_1}=180$ GeV. Colours refer to different values of $Q/$GeV: blue ($10^{3}$), red ($10^{7}$), green ($10^{10}$), purple ($10^{15}$) and cyan ($10^{19}$). The plots already encode our convention $m_{h_2} > m_{h_1}$ and the shaded red region refers to the values of $\alpha$ forbidden by LEP. Here: $x=3.5$ TeV, $m_{\nu_h}=200$ GeV.  \label{mh2_alpha}}
\end{figure}

Complementary to the previous study, we can now fix the light Higgs mass at specific, experimentally interesting\footnote{The chosen values maximise the probability for the decays $h_1\rightarrow b\overline{b}$, $h_1\rightarrow \gamma \gamma$, $h_1\rightarrow W^+W^-$ and $h_1\rightarrow ZZ$, respectively.}, values, i.e., $m_{h_1} = 100$, $120$, $160$ and $180$ GeV, and show the allowed region in the $m_{h_2}$ vs. $\alpha$ plane. This is done in figure~\ref{mh2_alpha}.

From this figures it is clear the transition of $h_2$ from the new extra scalar to the SM-like Higgs boson as we scan on the angle. As we increase $m_{h_1}$ (up to $m_{h_1} = 160$ GeV), a bigger region in $m_{h_2}$ is allowed for the model to be valid up to the Plank scale (the most inner regions, in cyan). Nonetheless, such a region exists also for a value of the light Higgs mass excluded by LEP for the SM, $m_{h_1} = 100$ GeV, but only for big values of the mixing angle. No new regions (with respect to the SM) in which the model can survive up to the Plank scale open for $m_{h_1} > 160$ GeV, as the allowed space deforms towards smaller values of $m_{h_1}$.

\subsection{Heavy neutrino mass influence}\label{sect:neutrino_eff}
\begin{figure}[!h]
  \subfloat[]{ 
  \label{mh1_mh2_a0_mhn-1000}
  \includegraphics[angle=0,width=0.48\textwidth ]{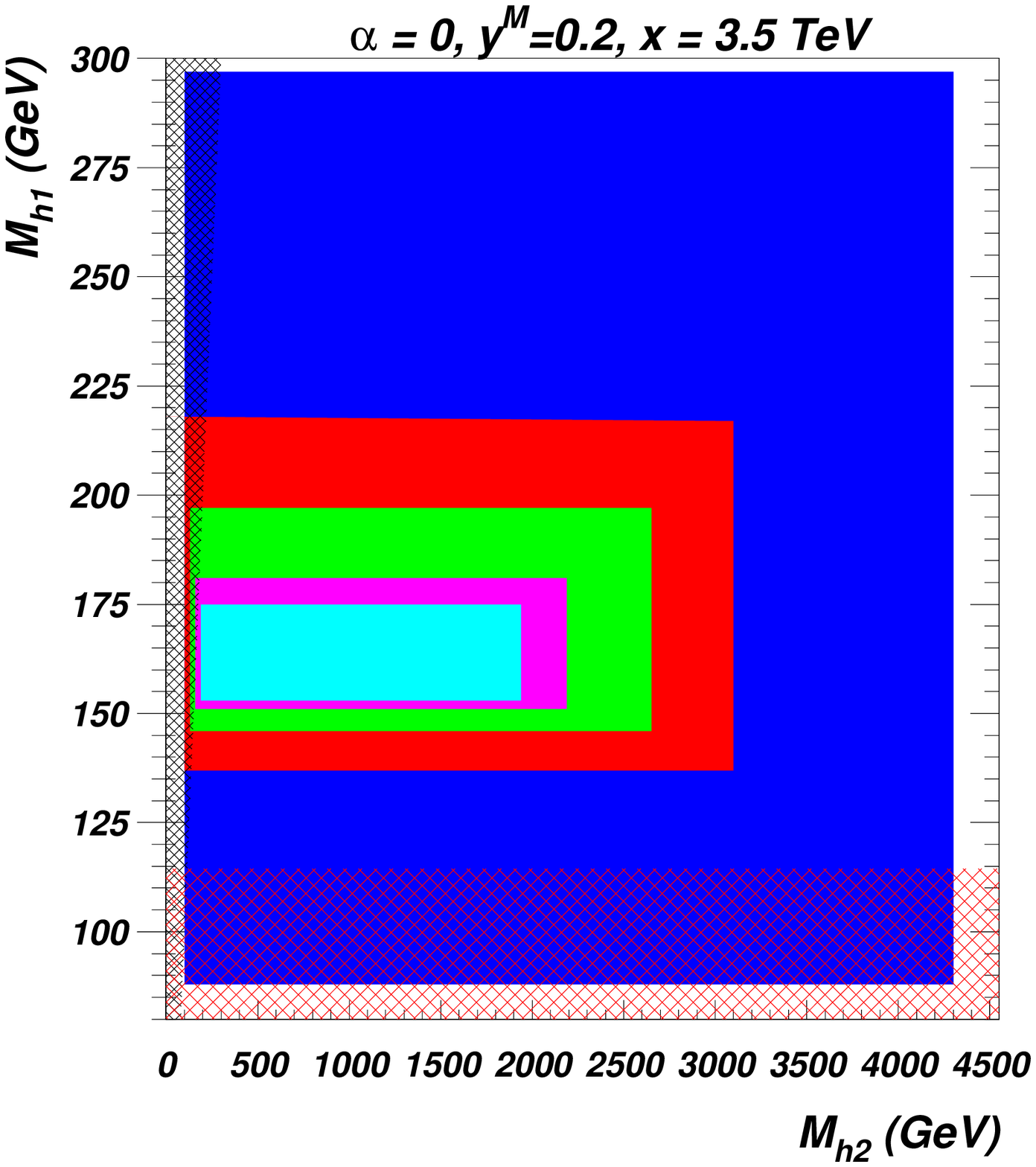}}
  \subfloat[]{
  \label{mh1_mh2_a01_mhn-1000}
  \includegraphics[angle=0,width=0.48\textwidth ]{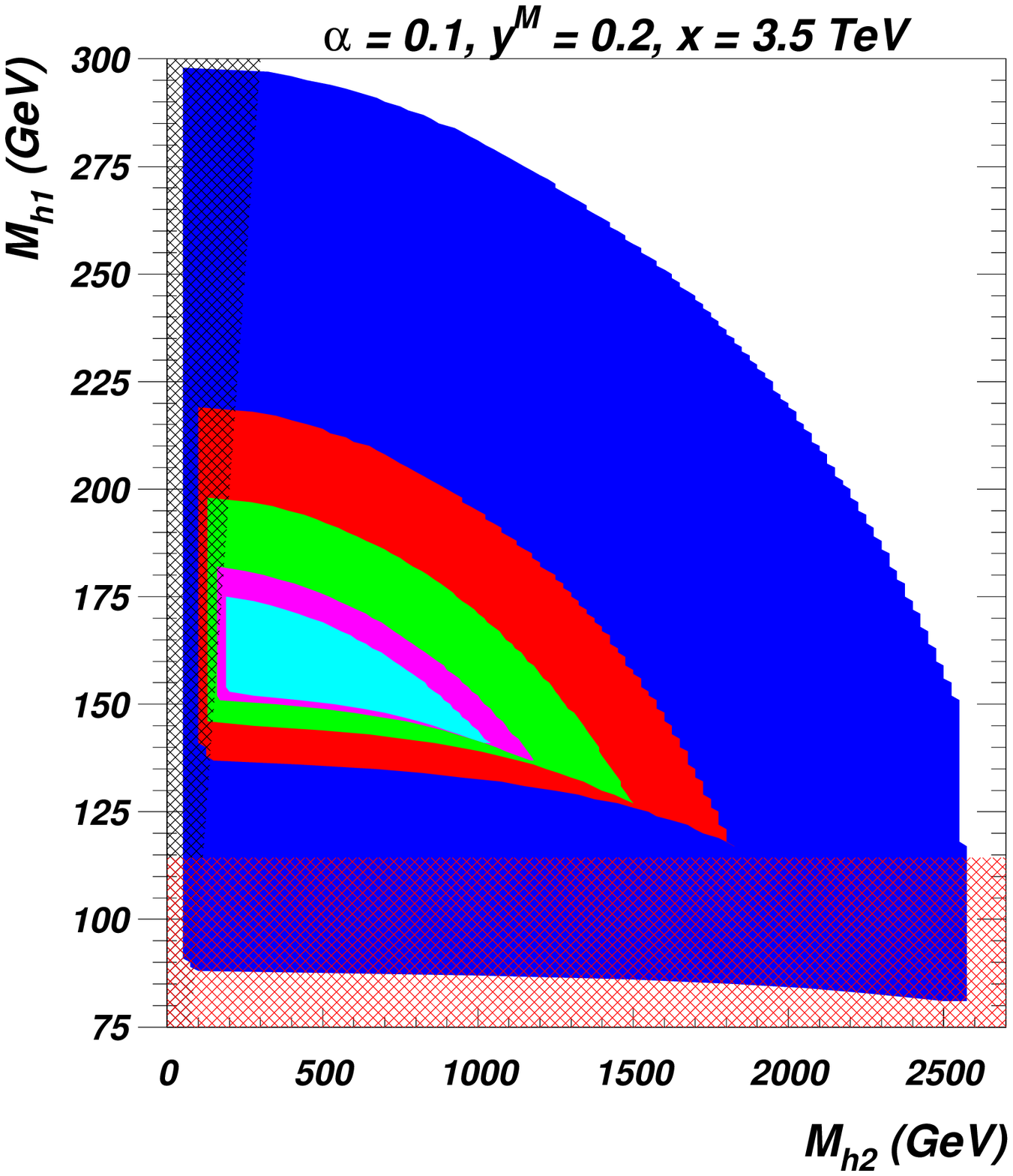}}
\\
  \subfloat[]{
  \label{mh1_mh2_api4_mhn-1000}
  \includegraphics[angle=0,width=0.48\textwidth ]{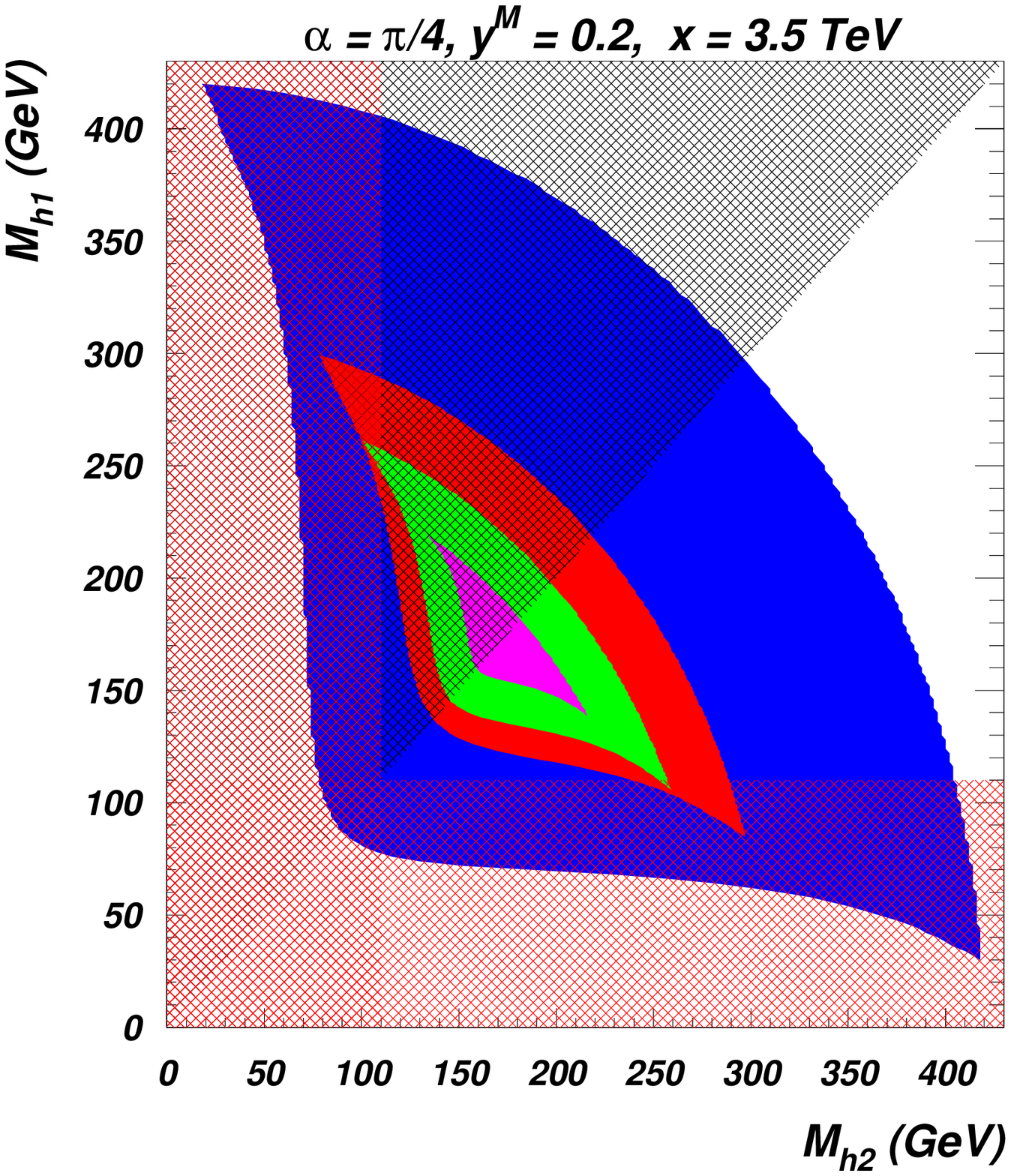}}
  \subfloat[]{
  \label{mh1_mh2_api3_mhn-1000}
  \includegraphics[angle=0,width=0.48\textwidth ]{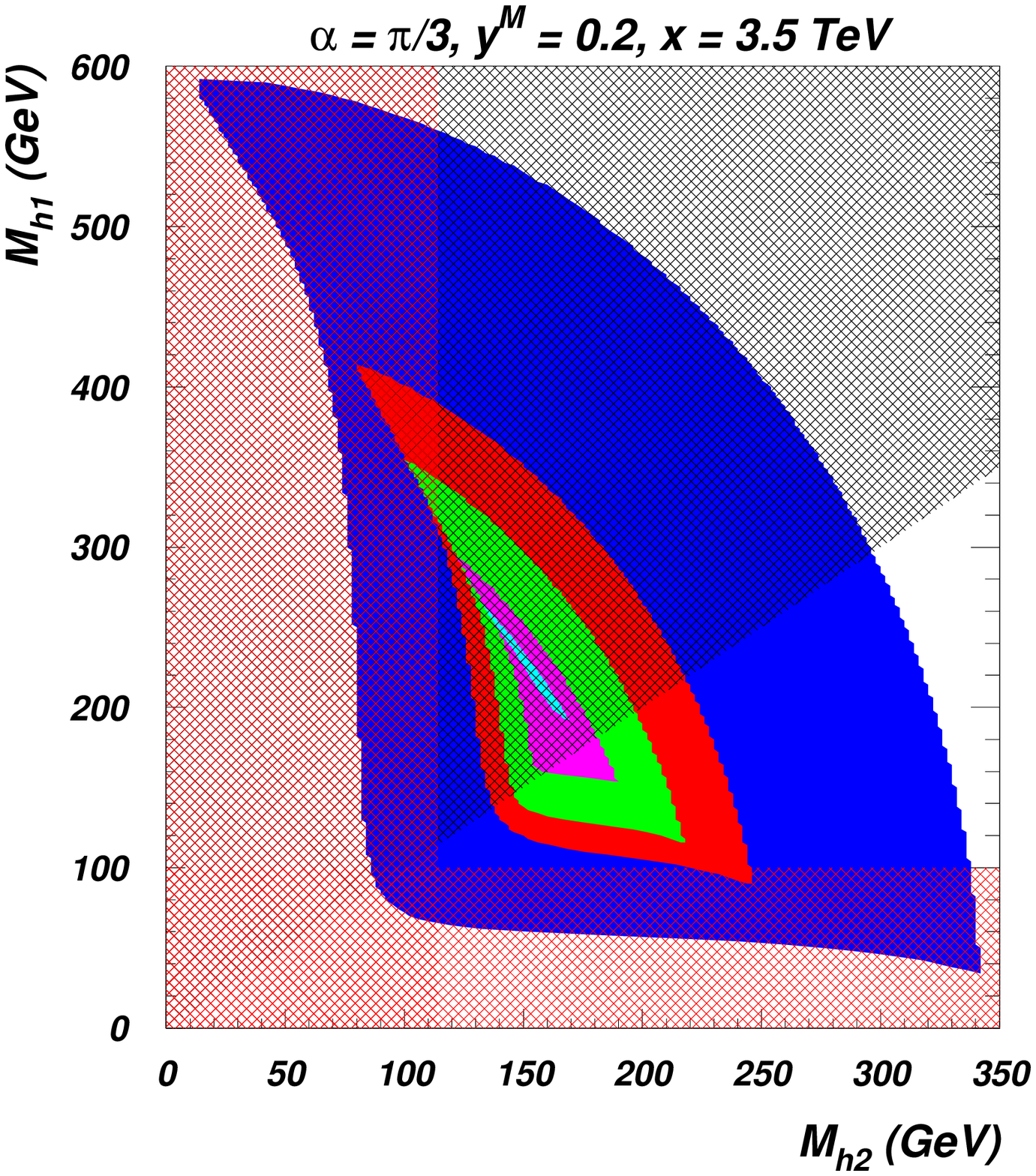}}
  \vspace*{-0.5cm}
  \caption{\it Allowed values in the $m_{h_1}$ vs. $m_{h_2}$ space  by eqs.~(\ref{cond_1}) and (\ref{cond_2}), for (\ref{mh1_mh2_a0_mhn-1000}) $\alpha =0$ and (\ref{mh1_mh2_a01_mhn-1000}) $\alpha =0.1$,  (\ref{mh1_mh2_api4_mhn-1000}) $\alpha = \pi /4$ and (\ref{mh1_mh2_api3_mhn-1000}) $\alpha = \pi /3$, for $m_{\nu_h}=1$ TeV and $x=3.5$ TeV. Colours refer to different values of $Q/$GeV: blue ($10^{3}$), red ($10^{7}$), green ($10^{10}$), purple ($10^{15}$) and cyan ($10^{19}$). The shaded black region is forbidden by our convention $m_{h_2} > m_{h_1}$, while the shaded red region refers to the values of the scalar masses forbidden by LEP.  \label{mh1_mh2_mhn}}
\end{figure}

As stated in section~\ref{sect:comp_det}, the RH neutrinos play for the extra scalar singlet the role of the top quark for the SM Higgs. This is particularly true for the vacuum stability condition, as the fermions in general provide the negative term that can drive the scalar couplings towards negative values. Figure~\ref{mh1_mh2_mhn} shows how the allowed regions in the $m_{h_1}$-$m_{h_2}$ plane change for a RH Majorana neutrino Yukawa coupling $y^M=0.2$ (that for $x=3.5$ TeV correspond to $m_{\nu_h} = 1$ TeV), not negligible anymore. For $y^M=0.4$, the changes are even more drastic, shrinking the allowed region even further.

The effect of having non negligible $y^M$ couplings is evident if we compare figure~\ref{mh1_mh2_mhn} to figure~\ref{mh1_mh2}. Notice that also the VEV $x$ is changed (from $7.5$ TeV to $3.5$ TeV), but this is only responsible for the smaller upper bounds of $m_{h_2}$ in figures~\ref{mh1_mh2_a0_mhn-1000} and \ref{mh1_mh2_a01_mhn-1000}. For small values of $\alpha$ it is evident our analogy between the top quark and the RH neutrinos, as now $m_{h_2}$ has a sensible lower bound too. The analogy holds also for bigger values of the angle, as the allowed region of masses is shrunk from below as we increase the RH Majorana neutrino Yukawa coupling, while the upper bound stays unaffected. The effect is even more evident for big values of the scale $Q$, with the Plank scale precluded now for whatever Higgs boson masses at $\alpha = \pi /4$ and tightly constraining the allowed ones at $\alpha = \pi /3$.

\begin{figure}[!h]
  \subfloat[]{ 
  \label{mh2_a_mh1-100_mhn1000}
  \includegraphics[angle=0,width=0.48\textwidth ]{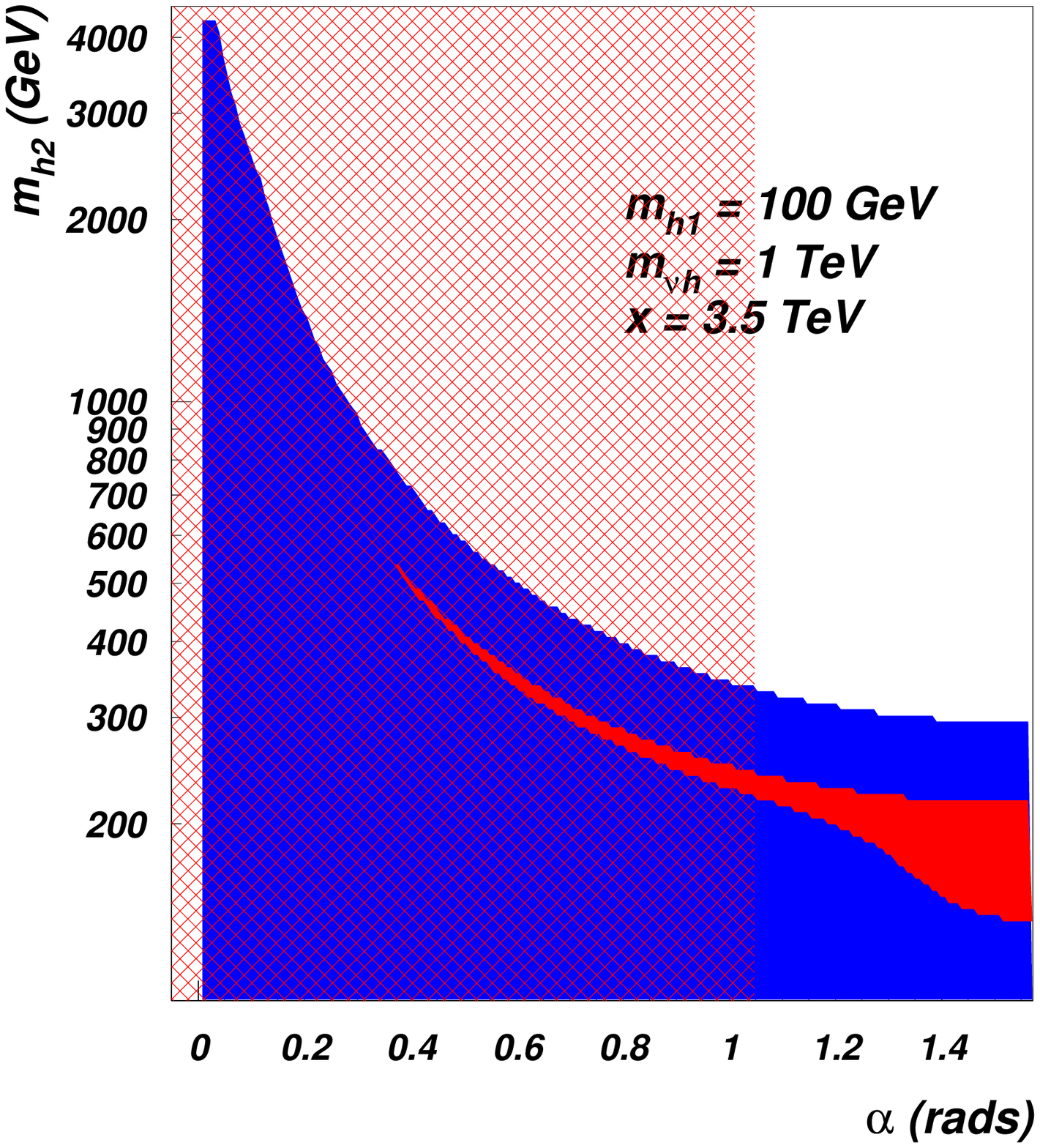}}
  \subfloat[]{
  \label{mh2_a_mh1-120_mhn1000}
  \includegraphics[angle=0,width=0.48\textwidth ]{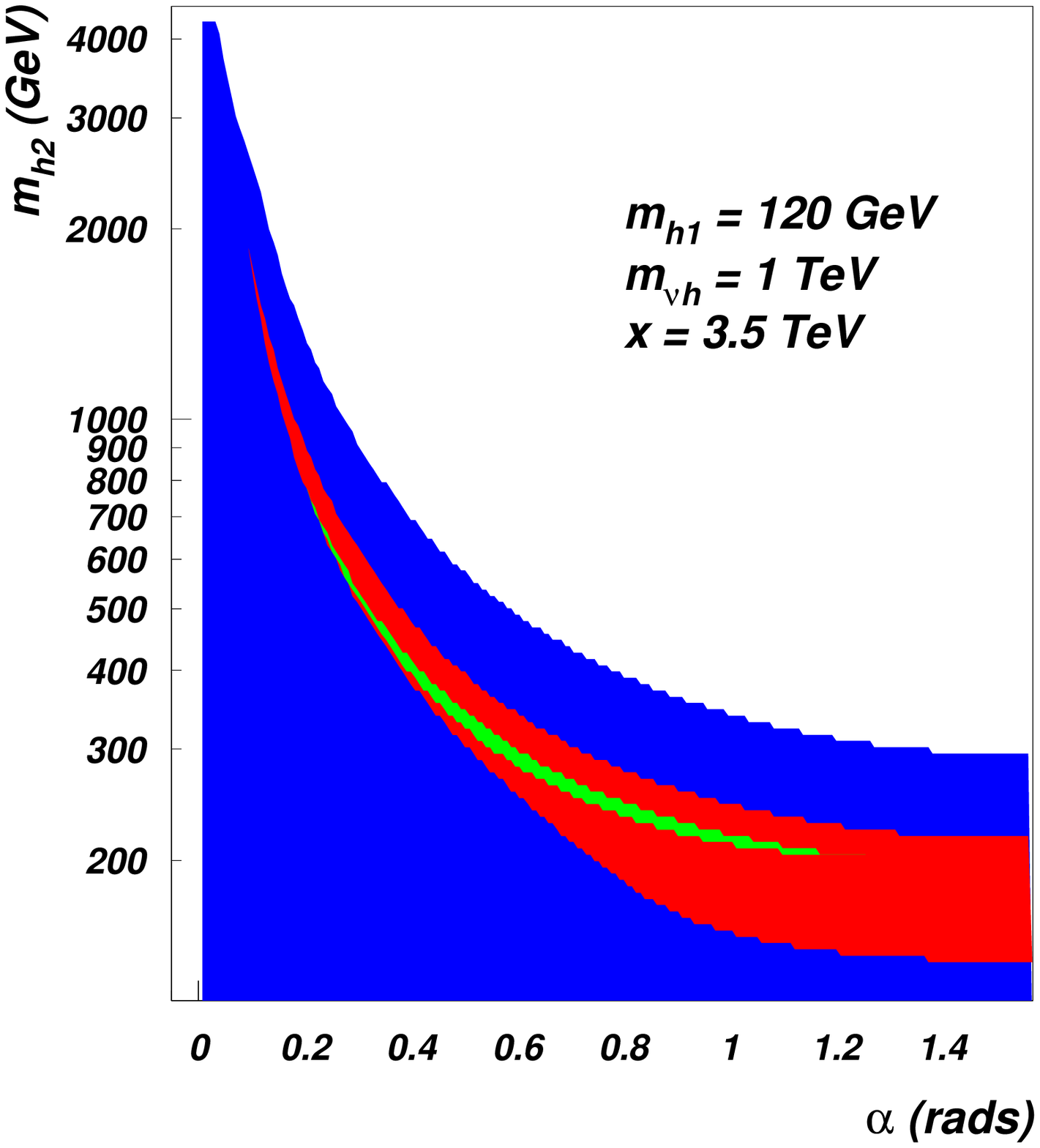}}
\\
  \subfloat[]{
  \label{mh2_a_mh1-160_mhn1000}
  \includegraphics[angle=0,width=0.48\textwidth ]{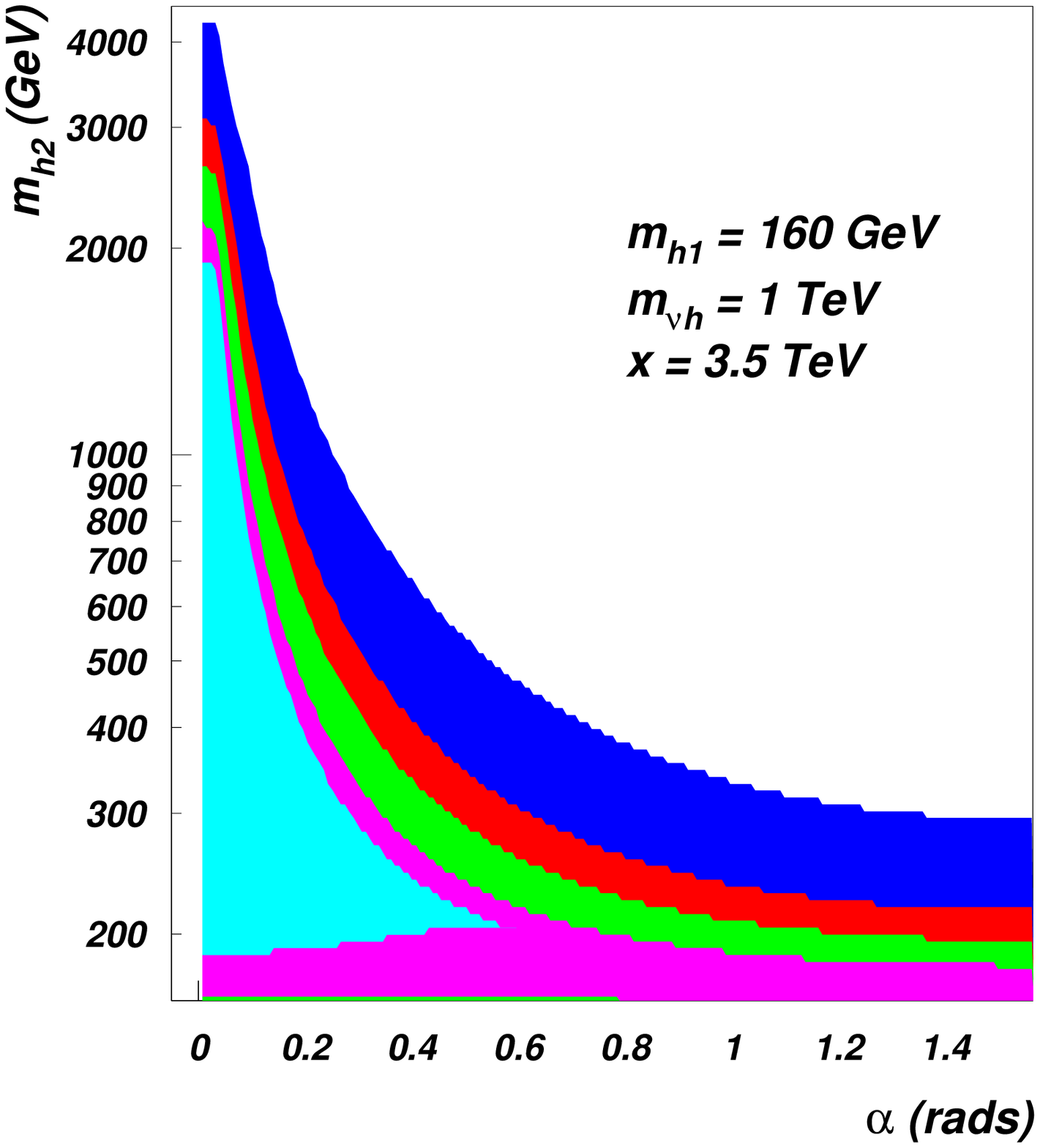}}
  \subfloat[]{
  \label{mh2_a_mh1-180_mhn1000}
  \includegraphics[angle=0,width=0.48\textwidth ]{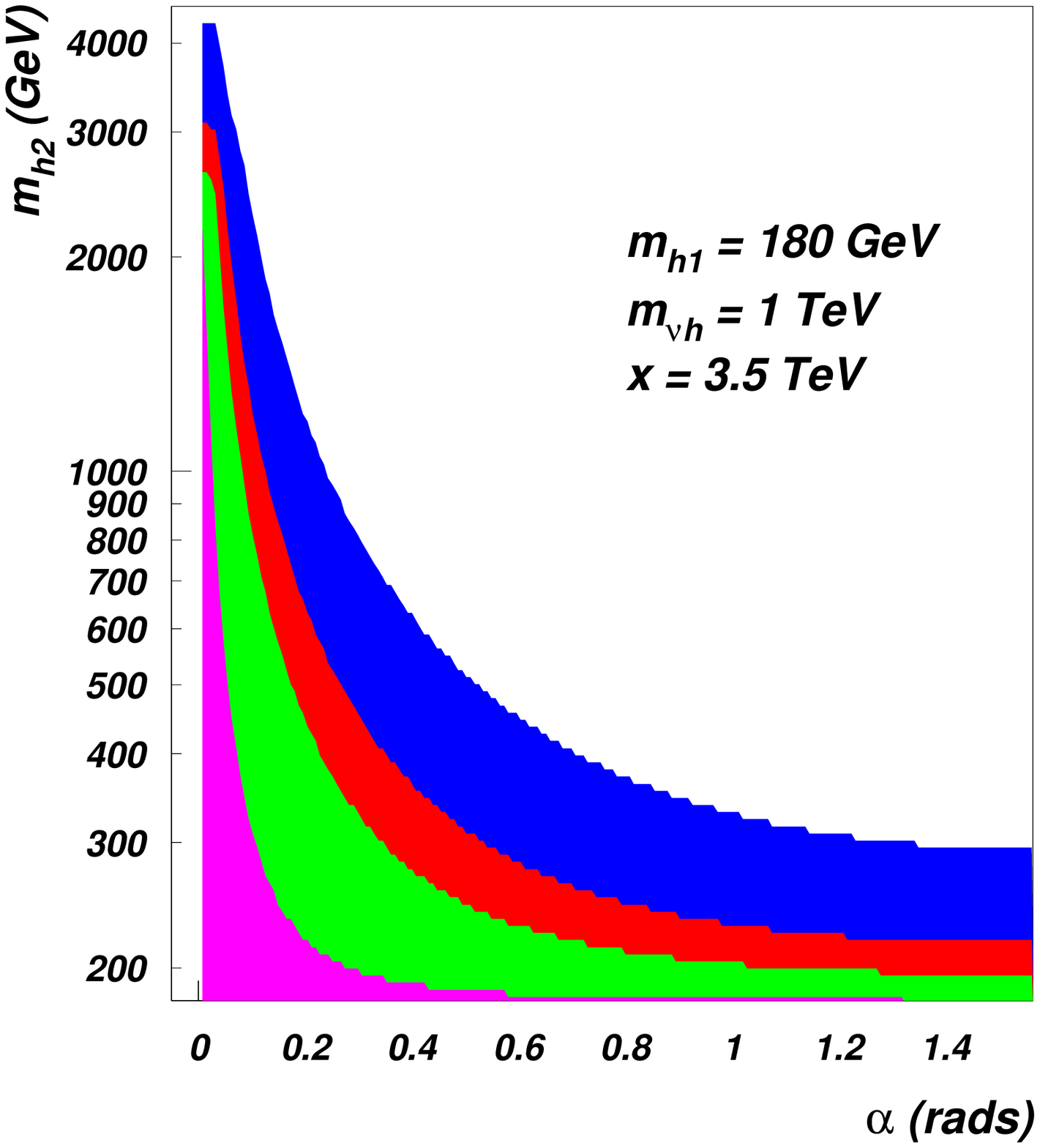}}
  \vspace*{-0.5cm}
  \caption{\it Allowed values in the $m_{h_2}$ vs. $\alpha$ space in the $B-L$ model by eqs.~(\ref{cond_1}) and (\ref{cond_2}), for (\ref{mh2_a_mh1-100_mhn1000}) $m_{h_1}=100$ GeV, (\ref{mh2_a_mh1-120_mhn1000}) $m_{h_1}=120$ GeV, (\ref{mh2_a_mh1-160_mhn1000}) $m_{h_1}=160$ GeV and (\ref{mh2_a_mh1-180_mhn1000}) $m_{h_1}=180$ GeV. Colours refer to different values of $Q/$GeV: blue ($10^{3}$), red ($10^{7}$), green ($10^{10}$), purple ($10^{15}$) and cyan ($10^{19}$). The plots already encode our convention $m_{h_2} > m_{h_1}$ and the shaded red region refers to the values of $\alpha$ forbidden by LEP. Here: $x=3.5$ TeV, $m_{\nu_h}=1$ TeV.  \label{mh2_alpha_mhn}}
\end{figure}

Moving to the $m_{h_2}$-$\alpha$ scan at fixed $m_{h_1}$ values, figure~\ref{mh2_alpha_mhn} shows the effect of the heavy neutrinos in this case, to be compared to figure~\ref{mh2_alpha}. It is evident that this model can survive until very large scales $Q$ with massive heavy neutrinos (for which, $y^M > 0.2$) only for the light Higgs boson masses allowed in the case of the SM, that is, $m_{h_1} \sim 160$ GeV. The mixing angle must also be small, $\alpha < \pi /5$, providing a tight constraint on $m_{h_2}$. For smaller $h_1$ masses, the effect of a large $y^M$ is to preclude scales $\displaystyle Q \gtrsim 10^7$ GeV almost completely, with for example just a tiny strip for $m_{h_1}=120$ GeV for which there exists a combination of $m_{h_2}$ and $\alpha$ such that the model is consistent up to $\displaystyle Q=10^{10}$ GeV. Finally, figure~\ref{mh2_a_mh1-180_mhn1000} is not visibly different from figure~\ref{mh2_a_mh1-180} just because we are showing only the $m_{h_2} > m_{h_1}$ region, the shrunk region being below.

\subsection{VEV effect}\label{sect:VEV_eff}
\begin{figure}[!h]
  \subfloat[]{
  \label{VEV_effect}
  \includegraphics[angle=0,width=0.48\textwidth ]{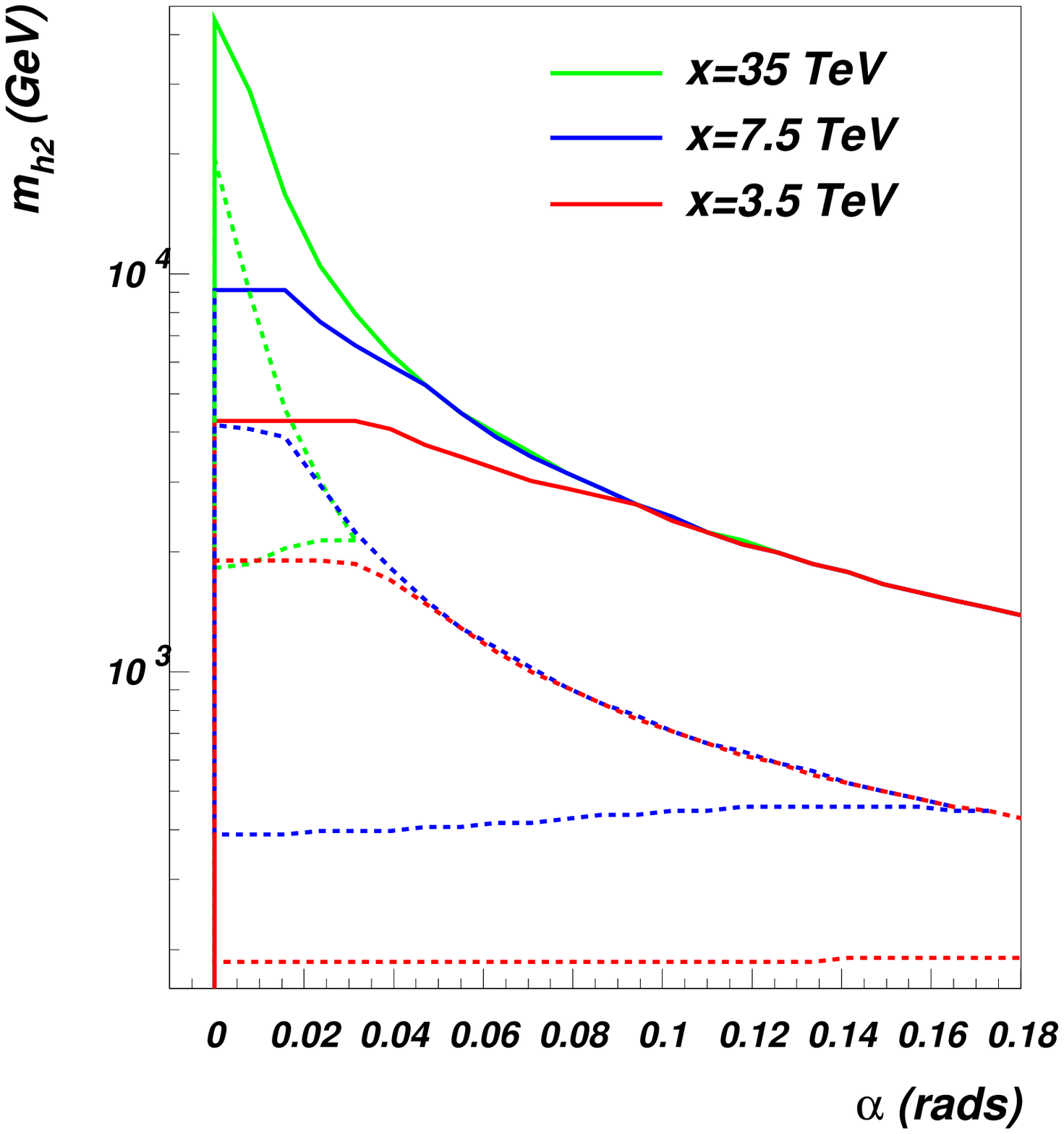}}
  \subfloat[]{
  \label{mh2_mh1_mhn1000_x75}
  \includegraphics[angle=0,width=0.48\textwidth ]{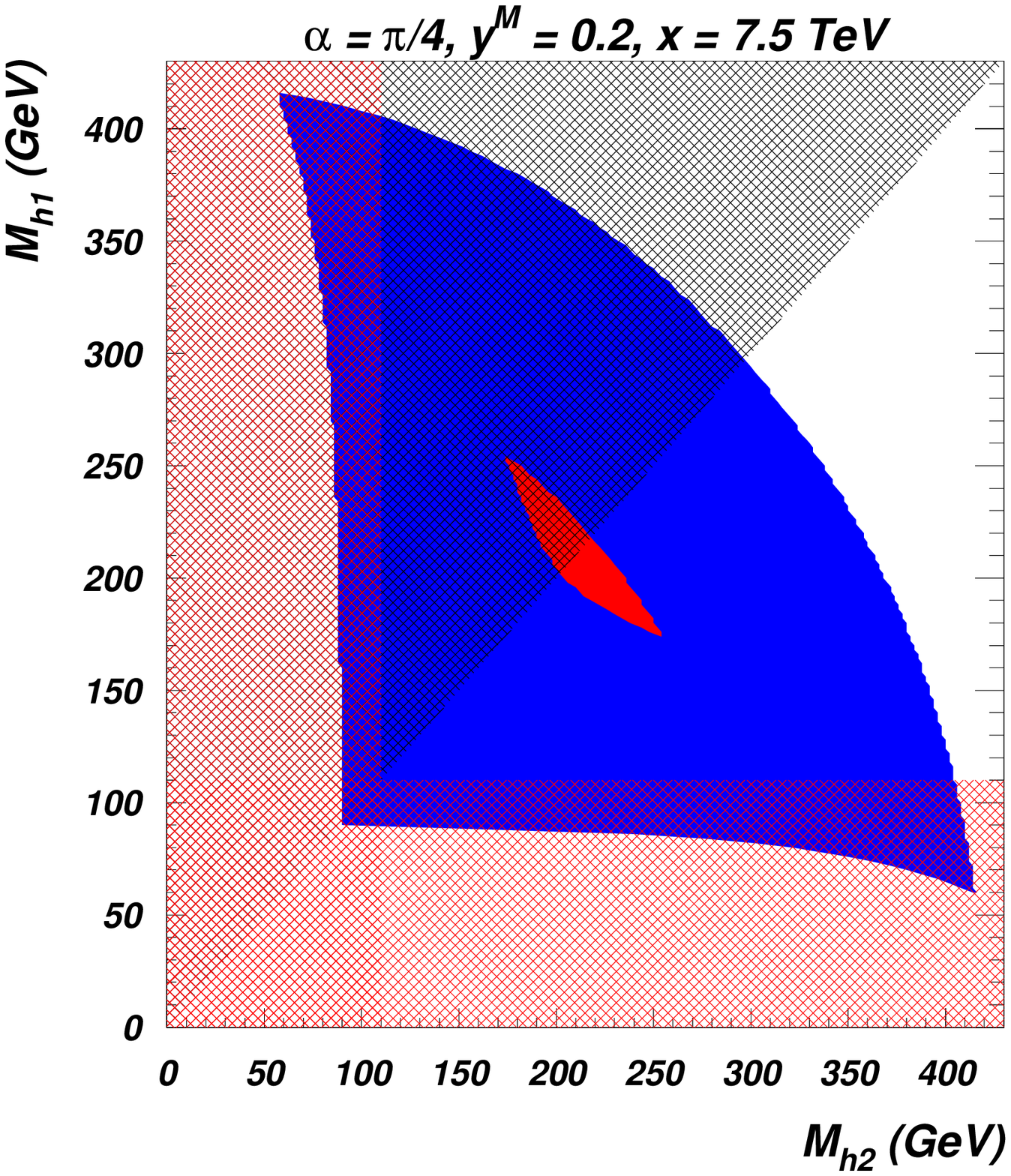}}
  \vspace*{-0.5cm}
  \caption{\it  Allowed values by eqs.~(\ref{cond_1}) and (\ref{cond_2}) (\ref{VEV_effect}) in the $m_{h_2}$ vs. $\alpha$ space for $m_{h_1}=160$ GeV and $y^M=0.2$, for $Q=10^3$ GeV (straight line) and $Q=10^{19}$ GeV (dashed line) for several $B-L$ breaking VEV values ($x=3.5$, $7.5$ and $35$ TeV, giving $m_{\nu _h}=1$, $2$ and $10$ TeV, respectively), and (\ref{mh2_mh1_mhn1000_x75}) in the $m_{h_1}$ vs. $m_{h_2}$ space, for $\alpha = \pi /4$, $x=7.5$ TeV and $y^M=0.2$, where colours refer to different values of $Q/$GeV: blue ($10^{3}$), red ($10^{7}$). The plots already encode our convention $m_{h_2} > m_{h_1}$ and the shaded red region refers to the values of $\alpha$ forbidden by LEP.  \label{mh2_mh1_mhn_api4}}
\end{figure}

The last effect to evaluate comes from changing the values for the $B-L$ breaking VEV $x$.
Figure~\ref{VEV_effect} shows the allowed regions in the $m_{h_2}$ vs. $\alpha$ plane for fixed $m_{h_1}=160$ GeV and $y^M = 0.2$ (that is, a particular case that shows all the interesting effects at once). As expected, since $\lambda _2$ is a function of $m_{h_2}/x$ (see for instance eq.~(\ref{inversion})), at $\alpha=0$ the bounds on $m_{h_2}$ simply scale linearly with the VEV. Regarding the upper bound, increasing the VEV $x$ naively increases the allowed region of the heavy Higgs masses, but it is remarkable that the effects are present only for small angles, $\alpha < 0.1$ radians, being the bigger angles unaffected. Concerning the lower bound, or the vacuum stability of the model, at fixed $y^M$, increasing the VEV $x$ requires to increase $m_{h_2}$ to keep $\lambda _2$ constant at the EW scale. This explains why, with non negligible $y^M$, the allowed heavy Higgs masses are shrinking from below when we increase the VEV $x$, as one can see in figure~\ref{VEV_effect} and comparing figure~\ref{mh2_mh1_mhn1000_x75} with figure~\ref{mh1_mh2_api4_mhn-1000}, both for $\alpha=\pi /4$ and $y^M=0.2$, but for $x=3.5$ and $x=7.5$ TeV, respectively.

In general, for the model to survive up to very large scales $Q\sim M_{\rm{Planck}}$, it is preferred the heavy neutrinos to be light with respect to the VEV $x$, in such a way that their Yukawa couplings are negligible in the RGE evolution of the scalar sector.

\section{Summary and conclusions}\label{sect:summa}
We have investigated the triviality and vacuum stability conditions of the minimal (or pure) $B-L$ model with a particular view to define the 
phenomenologically viable regions of the parameter space of the scalar sector, by computing all relevant RGEs (gauge, scalar and fermionic) at
the one-loop level in presence of all available experimental constraints. The RGE dependence on the Higgs masses and couplings (including mixings) 
has been studied in detail for selected heavy neutrino masses and couplings as well as discrete choices of the singlet Higgs field VEV.

Altogether, we have found that there exist configurations of the model for which its validity is guaranteed up to energy scales well beyond those
reachable at the LHC while at the same time enabling the CERN hadron collider to probe its scalar sector in Higgs mass and coupling regions completely
different from those accessible to the SM. Furthermore, we have shown that investigations of the Higgs sector of this extended scenario may also lead to 
constraints on other areas, such as the (heavy) neutrino and $Z'$ sectors (the latter indirectly, through the VEV of the singlet Higgs state
directly intervening in the scalar RGEs).

Combining the results of this paper on triviality and vacuum stability with those on unitarity of Ref.~\cite{B-L-unitarity}, we are now in a position
to investigate the production and decay phenomenology of both Higgs states of the minimal $B-L$ model at present and future accelerators \cite{newpap}.


\appendix
\section{RGEs}\label{RGs}
In this appendix we present the complete set of one-loop RGEs for the minimal $U(1)_{B-L}$ extension of the SM. For some parameters, the equations will be equal to those of the SM, as no extra contribution arises at one-loop level.

\subsection{Gauge RGEs}
The RGEs for the $SU(3)_C$ and $SU(2)_L$ gauge couplings $g_S$ and $g$ are \cite{Arason}:
\begin{eqnarray}\label{rge_gs}
\frac{d}{dt}g_S &=& \frac{g_S^3}{16\pi ^2}\left[ -11+\frac{4}{3}n_g \right] = \frac{g_S^3}{16\pi ^2}\left( -7\right)\, ,\\ \label{rge_g}
\frac{d}{dt}g &=& \frac{g^3}{16\pi ^2}\left[ -\frac{22}{3}+\frac{4}{3}n_g+\frac{1}{6}\right] = \frac{g^3}{16\pi ^2}\left(
		-\frac{19}{6}\right) \, ,
\end{eqnarray}
where $n_g =3$ is the number of generations.

Following standard techniques, we obtain for the Abelian couplings \cite{CPW,Aguila-Coughlan}:
\begin{eqnarray}\label{RGE_g1}
\frac{d}{dt}g_1 &=& \frac{1}{16\pi ^2}\left[A^{YY}g_1^3 \right]\, , \\ \label{RGE_g2}
\frac{d}{dt}g_1' &=& \frac{1}{16\pi ^2}\left[A^{XX}g_1'^3+2A^{XY}g_1'^2\widetilde g+A^{YY}g_1'\widetilde g^2 \right] \, , \\ \label{RGE_g_tilde}
\frac{d}{dt}\widetilde g &=& \frac{1}{16\pi ^2}\left[A^{YY}\widetilde{g}\,(\widetilde g^2+2g_1^2)+2A^{XY}g_1'(\widetilde{g}^2+g_1^2)+A^{XX}g_1'^2\widetilde g \right]\, ,
\end{eqnarray}
with
\begin{equation}\label{A_charge}
A^{ab} = A^{ba} = \frac{2}{3} \sum _f Q_f^a Q_f^b + \frac{1}{3}\sum _s Q_s^a Q_s^b\, , \qquad (a,b=Y,X)\, ,
\end{equation}
where the first sum is over the left-handed two-component fermions and the second one is over the complex scalars. For the model we are discussing ($Y$ is the SM weak hypercharge, $X=B-L$ is the $B-L$ number), the coefficients of eq.~(\ref{A_charge}) are, respectively:      
\begin{equation}
A^{YY}=41/6\, ,\qquad A^{XX}=12\, ,\qquad A^{YX}=16/3.
\end{equation}

\subsection{Fermion RGEs}
From straightforward calculations we obtain:
\begin{equation}\label{RGE_yuk_top}
\frac{d}{dt}y_t = \frac{y_t}{16\pi ^2}\left( \frac{9}{2}y_t^2-8g_S^2-\frac{9}{4}g^2-\frac{17}{12}g_1^2-\frac{17}{12}\widetilde{g}^2 -\frac{2}{3}g_1^{'2}-\frac{5}{3}\widetilde{g}g'_1 \right)\, .
\end{equation}
For the right-handed neutrinos, it is not restrictive to consider the basis in which the Majorana matrix of couplings is real, diagonal and positive:
$y^M \equiv \mbox{diag}\, (y^M_1,y^M_2,y^M_3)$. Then we get \cite{Luo_xiao,Iso:2009ss}\footnote{Notice the we get a difference of a factor $3$ in the third term in the RHS of the last expression in eq.~(14) contained in Ref.~\cite{Iso:2009ss}. The authors of Ref.~\cite{Iso:2009ss} acknowledged the difference and will correct their paper.}:
\begin{equation}\label{RGE_nu_r_maj}
\frac{d}{dt}y^M_i = \frac{y^M_i}{16\pi ^2}\left( 4(y^M_i)^2+2Tr\big[ (y^M)^2\big] -6g_1^{'2} \right)\, , \qquad (i=1\dots 3)\, .
\end{equation}

\subsection{Scalar RGEs}
A very straightforward way to find the one-loop RGEs for the parameters of the scalar potential is to compute the one-loop
effective potential and to impose its independence from the renormalisation scale. To one-loop level, the scalar potential $V$ reads:
\begin{equation}
V=V^{(0)}+\Delta V^{(1)}\, ,
\end{equation}
where $V^{(0)}$ is the tree-level potential and $\Delta V^{(1)}$ indicates the one-loop correction to it. To compute the latter it is useful to re-write the tree-level potential
\begin{equation}\label{new-potential}
V^{(0)}(H,\chi ) = m^2H^{\dagger}H + \mu ^2\mid\chi\mid ^2 + \lambda _1 (H^{\dagger}H)^2 +\lambda _2 \mid\chi\mid ^4 + \lambda _3 H^{\dagger}H\mid\chi\mid ^2 
\end{equation}
in terms of the real scalar fields:
\begin{equation}\label{H-CHI_degree}
H=\frac{1}{\sqrt{2}}\left( \begin{array}{c} \phi _1 +i\phi _2 \\ \phi _3 +i\phi _4\end{array}\right)\, , \hspace{2cm}
\chi =\frac{1}{\sqrt{2}}\left( \phi _5 +i\phi _6\right)\, .
\end{equation}
The only combinations of fields that are involved are $\phi ^2 =\phi _1^2+\phi _2^2+\phi _3^2+\phi _4^2 $ and
$\eta ^2 \equiv \phi _5^2+\phi _6^2$, so that eq.~(\ref{new-potential}) becomes:
\begin{equation}\label{tree_lev_pot}
V^{(0)}(\phi,\eta )= \frac{1}{2}m^2\phi ^2 +\frac{1}{2}\mu ^2\eta ^2 +\frac{1}{4}\lambda _1 \phi ^4 +\frac{1}{4}\lambda _2 \eta ^4 +
			\frac{1}{4}\lambda _3 \phi ^2 \eta ^2\, .
\end{equation}

The one-loop correction to the tree-level potential (\ref{tree_lev_pot}) is,
in the Landau gauge,
\begin{equation}\label{1-loop_pot}
\Delta V^{(1)}(\phi ,\eta )=\frac{1}{64\pi ^2}\sum _i(-1)^{2s_i}(2s_i+1)M^4_i(\phi ^2,\eta ^2)\left[ \ln{\frac{M^2_i(\phi ^2,\eta ^2)}{\mu ^2}-c_i}\right]\, ,
\end{equation}
where $c_i$ are constants that depend on the renormalisation scheme (for example, in the $\overline{\rm MS}$ scheme, it is $c_i=3/2$
for scalars and fermions, $c_i=5/6$ for vectors). Expanding eq.~(\ref{1-loop_pot}) and keeping the contributions of the scalar fields (Higgs
and Goldstone bosons), of the top-quark, of the gauge bosons and of the RH neutrinos only, we obtain
\begin{eqnarray*}
\Delta V^{(1)}&=&\frac{1}{64\pi ^2}\left\{ 3G_1^2\left[ \ln{\frac{G_1}{\mu ^2}-\frac{3}{2}}\right] + G_2^2\left[ \ln{\frac{G_2}{\mu ^2}-\frac{3}{2}}\right]
		+ Tr\left( H^2\left[ \ln{\frac{H}{\mu ^2}-\frac{3}{2}}\right]\right) \right. \\
	&&\left. -12T^2\left[ \ln{\frac{T}{\mu ^2}  -\frac{3}{2}}\right] + 3Tr\left( M_G^2\left[ \ln{\frac{M_G}{\mu ^2}-\frac{5}{6}}\right]\right) -2\sum _{i=1}^3 N_i^2\left[ \ln{\frac{N_i}{\mu ^2}-\frac{3}{2}}\right] \right\}\, ,
\end{eqnarray*}
where the field-dependent squared masses are, in a self-explanatory notation:
\begin{eqnarray}\label{3gold-field-dep}
G_1 (\phi ,\eta )&=& m^2+\lambda _1\phi ^2+\frac{\lambda _3}{2}\eta ^2\, ,\\ \label{1gold-field-dep}
G_2 (\phi ,\eta )&=& \mu ^2+\lambda _2\eta ^2+\frac{\lambda _3}{2}\phi ^2\, ,\\ \label{Higgs-field-dep}
H (\phi ,\eta )&=& \left( \begin{array}{cc} m^2+3\lambda _1\phi ^2+\frac{\lambda _3}{2}\eta ^2 & \lambda _3 \phi\eta\\ 
				\lambda _3\phi\eta & \mu ^2 +3\lambda _2\eta ^2+\frac{\lambda _3}{2}\phi ^2\end{array}\right)\, ,\\ \label{Top-field-dep}
T (\phi ,\eta )&=& \frac{1}{2}(y_t\phi )^2\, ,\\ \label{RH-N-field-dep}
M_G (\phi ,\eta ) &=& \frac{1}{4}\left(
		\begin{array}{ccc}
		g_1^{\phantom{o}2}\phi ^2 & -gg_1\phi ^2 & g_1\widetilde{g}\phi ^2\\
		-gg_1\phi ^2 & g^2\phi ^2 & -g\widetilde{g}\phi ^2\\
		g_1\widetilde{g}\phi ^2 & -g\widetilde{g}\phi ^2 & \widetilde{g}^2\phi ^2 + 16\eta ^2 g_1^{'2}
		\end{array} \right) \, ,\\ \label{Gauge_bosonos-dep}
N_i (\phi ,\eta )&=&\frac{1}{2}(y^M_i \eta )^2\, .
\end{eqnarray}
As usual, we define the beta functions $\beta _i$ ($i=1\dots 3$) for the quartic couplings, the gamma
functions $\gamma _{m,\mu}$ for the scalar masses and the scalar anomalous dimensions $\gamma _{\phi,\,\eta}$ as follows ($t=\ln{Q}$):
\begin{eqnarray}\label{beta_i}
\frac{d\lambda _i}{dt} &=& \beta _i\, ,\\ \label{gamma _m}
\frac{dm^2}{dt} &=& \gamma _m m^2\, ,\\ \label{gamma _mu}
\frac{d\mu ^2}{dt} &=& \gamma _\mu \mu ^2\, ,\\ \label{gamma _phi}
\frac{d\phi ^2}{dt} &=& 2\gamma _\phi \phi^2\, ,\\ \label{gamma _eta}
\frac{d\eta ^2}{dt} &=& 2\gamma _\eta \eta ^2\, .
\end{eqnarray}

Now we can extract the RGEs for the parameters of the scalar potential just by requiring that the first derivative of the effective
potential with respect to the scale $t$ vanishes:
\begin{equation}\label{eff_pot_der_nulla}
\frac{d}{dt}V^{(1)} \equiv \frac{d}{dt}(V^{(0)}+\Delta V^{(1)})\equiv 0\, ,
\end{equation}
keeping only the one-loop terms. Reorganising it in a more convenient way, we see that eq.~(\ref{eff_pot_der_nulla}) implies the following
equations: 
\begin{eqnarray*}\label{RGE_1}
\frac{m^2\phi ^2}{2} \left[ \gamma _m +2\gamma _\phi  -\frac{1}{16\pi ^2}\left( 12\lambda _1+2\frac{\mu ^2}{m^2}\lambda _3\right) \right] &=& 0\, ,\\ \label{RGE_2}
\frac{\mu ^2 \eta ^2}{2} \left[\gamma _\mu +2\gamma _\eta -\frac{1}{16\pi ^2}\left( 8\lambda _2+4\frac{m^2}{\mu ^2}\lambda _3\right) \right] &=& 0\, ,\\ \label{RGE_3}
\frac{\phi ^4}{4}\left[\beta _1 +4\lambda _1\gamma _\phi -\frac{1}{16\pi ^2}\left( 24\lambda _1^2+\lambda _3^2
	-6y_t^4
	+\frac{9}{8}g^4+\frac{3}{8}g_1^4+\frac{3}{4}g^2g_1^2 \right.\right. \qquad && \\
	 \left.\left.
	+\frac{3}{4}g^2\widetilde{g}^2+\frac{3}{4}g_1^2\widetilde{g}^2 +\frac{3}{8}\widetilde{g}^4 \right)\right] &=&0\, ,\\ \label{RGE_4}
\frac{\eta ^4}{4}\left[\beta _2 +4\lambda _2\gamma _\eta -\frac{1}{8\pi ^2}\left( 10\lambda _2^2+\lambda _3^2
	-\frac{1}{2}Tr\big[ (y^M)^4\big] +48 g_1^{'4}\right)\right] &=&0\, ,\\ \label{RGE_5}
\frac{\phi ^2\eta ^2}{4} \left[\beta _3 +2\lambda _3(\gamma _\phi  +\gamma _\eta )-\frac{1}{8\pi ^2}\left( 6\lambda _1\lambda _3+4\lambda _2\lambda _3 +2\lambda _3^2 +6\widetilde{g}^2 g_1^{'2} \right)\right] &=&0\, .
\end{eqnarray*}

Imposing that each term between squared brackets vanishes, we can obtain the RGEs for the parameters of the scalar potential after
inserting the explicit expression of the scalar anomalous dimensions $\gamma _\phi$ and $\gamma _\eta$. The latter are easily
computed and read \cite{Luo_xiao,Iso:2009ss,sher}:
\begin{eqnarray}\label{gamma_phi}
\gamma _\phi &=& -\frac{1}{16 \pi ^2} \left( 3y_t^2 -\frac{9}{4}g^2-\frac{3}{4}g_1^2-\frac{3}{4}\widetilde{g}^2 \right)\, ,\\ \label{gamma_eta}
\gamma _\eta &=& -\frac{1}{16 \pi ^2} \left( 2Tr\left[ (y^M)^2\right] - 12 g_1^{'2}\right)\, .
\end{eqnarray}
Inserting eqs. (\ref{gamma_phi}) and (\ref{gamma_eta}) into the RGEs, we finally obtain the RGEs for the five parameters in the scalar potential:
\begin{eqnarray}\label{RGE_m}
\gamma _m \equiv \frac{1}{m^2}\frac{d m^2}{dt} &=& \frac{1}{16\pi ^2}\left( 12\lambda _1 +6y_t^2+2\frac{\mu ^2}{m^2}\lambda _3 -\frac{9}{2}g^2-\frac{3}{2}g_1^2-\frac{3}{2}\widetilde{g}^2\right)\, , \\ \label{RGE_mu}
\gamma _\mu \equiv \frac{1}{\mu ^2}\frac{d \mu ^2}{dt} &=& \frac{1}{16\pi ^2}\left( 8\lambda _2+4Tr\left[ (y^M)^2\right] +4\frac{m^2}{\mu ^2}\lambda _3 - 24 g_1^{'2}\right)\, ,\\ \nonumber
\beta _1 \equiv \frac{d \lambda _1}{dt} &=&  \frac{1}{16\pi ^2}\left( 24\lambda _1^2+\lambda _3^2
-6y_t^4 +\frac{9}{8}g^4 +\frac{3}{8}g_1^4 +\frac{3}{4}g^2g_1^2 +\frac{3}{4}g^2\widetilde{g}^2   \right. \\ \label{RGE_lamda1}
&& \left. +\frac{3}{4}g_1^2\widetilde{g}^2+\frac{3}{8}\widetilde{g}^4 + 12\lambda _1 y_t^2 -9\lambda _1 g^2-3\lambda _1 g_1^2-3\lambda _1 \widetilde{g}^2
	\right)\, ,\\ \nonumber
\beta _2 \equiv \frac{d \lambda _2}{dt} &=&  \frac{1}{8\pi ^2}\left( 10\lambda _2^2+\lambda _3^2-\frac{1}{2}Tr\left[ (y^M)^4\right] +48 g_1^{'4}+4\lambda _2Tr\left[ (y^M)^2\right] \right.\\ \label{RGE_lamda2}
&& \left. -24\lambda _2g_1^{'2} \right)\, ,\\ \nonumber
\beta _3 \equiv \frac{d \lambda _3}{dt} &=&  \frac{\lambda _3}{8\pi ^2}\left( 6\lambda _1+4\lambda _2+2\lambda _3+3y_t^2-\frac{9}{4}g^2-\frac{3}{4}g_1^2-\frac{3}{4}\widetilde{g}^2 \right.  \\ \label{RGE_lamda3}
&& \left. +2Tr\left[ (y^M)^2\right] - 12 g_1^{'2} + 6\frac{\widetilde{g}^2 g_1^{'2}}{\lambda _3}\right)\, .
\end{eqnarray}

\section*{Acknowledgements}\label{acknowledgements}
We would like to thank A. Belyaev for most useful comments and discussions. The work of all of us is supported
in part by the NExT Institute.



\end{document}